\documentclass{IEEEtran}

\usepackage{amssymb,amsxtra,amsmath,latexsym,graphics,verbatim,
epsfig,setspace,colortbl,nicefrac,multirow,algorithm,algorithmic,theorem,times,
subfigure,mathrsfs}

\renewenvironment{IEEEproof}[1]{\noindent {\it Proof~:} #1}
{\ \rule{1mm}{2mm}}
\theorembodyfont{\normalfont\slshape}

\newtheorem{theorem}{Theorem}
\newtheorem{lemma}[theorem]{Lemma}
\newtheorem{proposition}[theorem]{Proposition}
\newtheorem{corollary}[theorem]{Corollary}

\def\b{\beta}

\def\e{\varepsilon}
\def\g{{\qopname\relax{no}{g}}}

\def\b{{\boldsymbol b}}
\def\c{{\boldsymbol c}}
\def\s{{\boldsymbol s}}
\def\u{{\boldsymbol u}}

\def\c{{\boldsymbol c}}
\def\b{{\boldsymbol b}}
\def\c{{\boldsymbol c}}
\def\s{{\boldsymbol s}}
\def\u{{\boldsymbol u}}
\def\e{{\boldsymbol e}}
\def\h{{\boldsymbol h}}
\def\x{{\boldsymbol x}}
\def\y{{\boldsymbol y}}
\def\z{{\boldsymbol z}}
\def\0{{\boldsymbol 0}}
\def\1{{\boldsymbol 1}}
\newcommand\nc\newcommand
\nc\bfa{{\boldsymbol a}}\nc\bfA{{\bf A}}\nc\cA{{\mathcal A}}\nc\sA{{\mathscr A}}
\nc\bfb{{\boldsymbol b}}\nc\bfB{{\bf B}}\nc\cB{{\mathcal B}}\nc\sB{{\mathscr B}}
\nc\bfc{{\boldsymbol c}}\nc\bfC{{\bf C}}\nc\cC{{\mathcal C}}\nc\sC{{\mathscr C}}
\nc\bfd{{\boldsymbol d}}\nc\bfD{{\bf D}}\nc\cD{{\mathcal D}}\nc\sD{{\mathscr D}}
\nc\bfe{{\boldsymbol e}}\nc\bfE{{\bf E}}\nc\cE{{\mathcal E}}\nc\sE{{\mathscr E}}
\nc\bff{{\boldsymbol f}}\nc\bfF{{\bf F}}\nc\cF{{\mathcal F}}\nc\sF{{\mathscr F}}
\nc\bfg{{\boldsymbol g}}\nc\bfG{{\bf G}}\nc\cG{{\mathcal G}}\nc\sG{{\mathscr G}}
\nc\bfh{{\boldsymbol h}}\nc\bfH{{\bf H}}\nc\cH{{\mathcal H}}\nc\sH{{\mathscr H}}
\nc\bfi{{\boldsymbol i}}\nc\bfI{{\bf I}}\nc\cI{{\mathcal I}}\nc\sI{{\mathscr I}}
\nc\bfj{{\boldsymbol j}}\nc\bfJ{{\bf J}}\nc\cJ{{\mathcal J}}\nc\sJ{{\mathscr J}}
\nc\bfk{{\boldsymbol k}}\nc\bfK{{\bf K}}\nc\cK{{\mathcal K}}\nc\sK{{\mathscr K}}
\nc\bfl{{\boldsymbol l}}\nc\bfL{{\bf L}}\nc\cL{{\mathcal L}}\nc\sL{{\mathscr L}}
\nc\bfm{{\boldsymbol m}}\nc\bfM{{\bf M}}\nc\cM{{\mathcal M}}\nc\sM{{\mathscr M}}
\nc\bfn{{\boldsymbol n}}\nc\bfN{{\bf N}}\nc\cN{{\mathcal N}}\nc\sN{{\mathscr N}}
\nc\bfo{{\boldsymbol o}}\nc\bfO{{\bf O}}\nc\cO{{\mathcal O}}\nc\sO{{\mathscr O}}
\nc\bfp{{\boldsymbol p}}\nc\bfP{{\bf P}}\nc\cP{{\mathcal P}}\nc\sP{{\mathscr P}}
\nc\bfq{{\boldsymbol q}}\nc\bfQ{{\bf Q}}\nc\cQ{{\mathcal Q}}\nc\sQ{{\mathscr Q}}
\nc\bfr{{\boldsymbol r}}\nc\bfR{{\bf R}}\nc\cR{{\mathcal R}}\nc\sR{{\mathscr R}}
\nc\bfs{{\boldsymbol s}}\nc\bfS{{\bf S}}\nc\cS{{\mathcal S}}\nc\sS{{\mathscr S}}
\nc\bft{{\boldsymbol t}}\nc\bfT{{\bf T}}\nc\cT{{\mathcal T}}\nc\sT{{\mathscr T}}
\nc\bfu{{\boldsymbol u}}\nc\bfU{{\bf U}}\nc\cU{{\mathcal U}}\nc\sU{{\mathscr U}}
\nc\bfv{{\boldsymbol v}}\nc\bfV{{\bf V}}\nc\cV{{\mathcal V}}\nc\sV{{\mathscr V}}
\nc\bfw{{\boldsymbol w}}\nc\bfW{{\bf W}}\nc\cW{{\mathcal W}}\nc\sW{{\mathscr W}}
\nc\bfx{{\boldsymbol x}}\nc\bfX{{\bf Z}}\nc\cX{{\mathcal X}}\nc\sX{{\mathscr X}}
\nc\bfy{{\boldsymbol y}}\nc\bfY{{\bf Y}}\nc\cY{{\mathcal Y}}\nc\sY{{\mathscr Y}}
\nc\bfz{{\boldsymbol z}}\nc\bfZ{{\bf Z}}\nc\cZ{{\mathcal Z}}\nc\sZ{{\mathscr Z}}
\nc\half{\nicefrac12}
\def\eps{\epsilon}\def\hc{\hat{c}}

\nc{\remove}[1]{}
\nc\nd\noindent

\def\hc{\hat{c}}
\def\hbc{\hat{\c}}

\def\oR{\overline{R}}
\def\uR{\underline{R}}

\def\wt{{w_H}}

\def\dist{{d_H}}
\def\h{h}
\def\Pr{\qopname\relax{no}{Pr}}

\def\H{\qopname\relax{no}{H}}

\begin{document}
\thispagestyle{empty}

\title{Coding for High-Density Recording on a 1-D Granular Magnetic Medium}
\author{Arya Mazumdar, Alexander Barg, Navin Kashyap

\thanks{Arya Mazumdar is with the Department of
    Electrical and Computer Engineering and Institute for Systems
    Research, University of Maryland, College Park, MD 20742 (e-mail:
    arya@umd.edu).}
    \thanks{Alexander Barg is with the Department of Electrical
    and Computer Engineering and Institute for Systems Research,
    University of Maryland, College Park, MD 20742 and Institute for
    Problems of Information Transmission, Moscow, Russia (e-mail:
    abarg@umd.edu).}
    \thanks{Navin Kashyap is with the Department of Mathematics and Statistics,
Queen's University, Kingston, ON K7L3N6, Canada (e-mail:
    nkashyap@mast.queensu.ca).}
    \thanks{The work of Arya Mazumdar and Alexander Barg was supported
by NSF grants CCF0830699 and CCF0916919. The work of Navin Kashyap
was supported in part by a Discovery Grant from NSERC, Canada,
and was performed while on sabbatical at the University of Maryland,
College Park, and the Indian Institute of Science, Bangalore.

The results of this paper were presented in part at the 2010 IEEE International
Symposium on Information Theory, Austin, TX, June 13-18, 2010, and at the
48th Annual Allerton Conference on Communication, Control, and Computing, 
Monticello, IL, September 29--October 1, 2010.} }

\maketitle

\begin{abstract}
In terabit-density magnetic recording, several bits of data
can be replaced by the values of their neighbors in the
storage medium. As a result, errors in the medium are
dependent on each other and also on the data written.
We consider a simple one-dimensional combinatorial model of this medium.
In our model, we assume a setting where binary data is sequentially
written on the medium and a bit can erroneously change to
the immediately preceding value. We derive several properties of
 codes that correct this type of errors, focusing on bounds on 
their cardinality.

We also define a probabilistic finite-state channel model of the 
storage medium, and derive lower and upper estimates of its capacity. 
A lower bound is derived by evaluating the symmetric capacity of the
channel, i.e., the maximum transmission rate under the assumption
of the uniform input distribution of the channel.
An upper bound is found by showing that the original channel is
a stochastic degradation of another, related channel model
whose capacity we can compute explicitly.
\end{abstract}

\section{Introduction}

One of the challenges in achieving ultra-high-density magnetic recording
lies in accounting for the effect of the granularity of the
recording medium. Conventional magnetic recording media are
composed of fundamental magnetizable units, called ``grains'',
that do not have a fixed size or shape. Information is stored on
the medium through a write mechanism that sets the magnetic polarities
of the grains \cite{whi97}. There are two types of magnetic polarity,
and each grain can be magnetized to take on exactly one
of these two polarities. Thus, each grain can store at most
one bit of information. Clearly, if the boundaries
of the grains were known to the write mechanism
and the readback mechanism, then it would be theoretically possible
to achieve a storage capacity of one information bit per grain.

There are two bottlenecks to achieving the one-bit-per-grain storage capacity:
(i)~the existing write (and readback) technologies are not capable of
setting (and reading back) the magnetic polarities of a region as small
as a single grain; and (ii)~the write and readback mechanisms
are typically unaware of the shapes and positions of the grains
in the medium. In current magnetic recording technologies, writing is
generally done by dividing the magnetic medium into regularly-spaced bit cells,
and writing one bit of data into each of these bit cells.
The bit cells are much larger in size compared to the grains,
so that each bit cell comprises many grains.
Writing a bit into a bit cell is then a matter of uniformly magnetizing
all the grains within the cell; the effect of grains straddling
the boundary between two bit cells can be neglected.

Recently, Wood et al.~\cite{WWKM09} proposed a new write mechanism,
that can magnetize areas commensurate to the size of individual grains.
With such a write mechanism and a corresponding readback mechanism in place,
the remaining bottleneck to achieving magnetic recording densities as high
as 10 Terabits per square inch is that the
write and readback mechanisms do not have precise knowledge of
the grain boundaries.

The authors of \cite{WWKM09} went on to consider the information loss caused
by the lack of knowledge of grain boundaries. A sample simulation
considered a two-dimensional magnetic medium composed of 100
randomly shaped grains, and subdivided into a $14 \times 14$
grid of uniformly-sized bit cells. Bits were written in raster-scan fashion
onto the grid. At the $k$th step of the write process, if any grain
had more than a 30\% (in area) overlap with the bit cell to be written
at that step, then that grain was given the polarity value of the $k$th bit.
The polarity of a grain could switch multiple times before settling
on a final value. With a readback mechanism that reported the
polarity value at the centre of each bit cell, their simulation
recorded the proportion of bits that were reported with the wrong polarity.
A similar simulation, but with a slightly different assumption on the
underlying grain distribution, was reported in \cite{KRV09}.

The authors of \cite{WWKM09} also considered a simple channel
that modeled a one-dimensional granular medium, and computed a lower
bound on the capacity of the channel. The one-dimensional medium
was divided into regularly-spaced bit cells, and it was assumed that
grain boundaries coincided with bit cell boundaries, and that the
grains had randomly selected lengths equal to 1, 2 or 3 bit cells.
The polarity of a grain is set by the last bit to be written within it.
The effect of this is that the last bit to be written in the grain
\emph{overwrites} all bits previously written within the same grain.

In this paper, we restrict ourselves to the one-dimensional case,
and consider a combinatorial error model that corresponds to the
granular medium described above. The medium comprises $n$ bit cells,
indexed by the integers from 1 to $n$. The granular structure of the medium
is described by an increasing sequence of positive integers,
$1 = j_1 < j_2 < \cdots < j_s \leq n$, where $j_i$ denotes the
index of the bit cell at which the $i$th grain begins.
Note that the length of the $i$th grain is $\ell_i = j_{i+1} - j_i$
(we set $j_{s+1} = n+1$ to be consistent).

The effect of a given grain pattern on an $n$-bit block of binary data
$\bfx=(x_1,x_2,\dots,x_n)$ to be written onto the medium is
represented by an operator $\phi$ that acts upon $\bfx$ to produce
$\phi(\bfx) = (y_1, y_2, \ldots, y_n)$, which is the binary vector that
is actually recorded on the medium. For notational ease, our model assumes
that it is the \emph{first} bit to be written within a grain that sets the
polarity of the grain. Thus, for indices $j$ within the $i$th grain,
\emph{i.e.}, for $j_i \leq j < j_{i+1}$, we have $y_j = x_{j_i}$.
This means that the $i$th grain introduces an error in the recorded data
(\emph{i.e.}, a situation where $y_j \neq x_j$) precisely when
$x_j \neq x_{j_i}$ for some $j$ satisfying $j_i < j < j_{i+1}$.
In particular, grains of length 1 do not introduce any errors.

As an example, consider a medium divided into 15 bit cells,
with a granular structure consisting of grains of lengths 1 and 2 only,
with the length-2 grains beginning at indices 3, 6, 8 and 13.
The grains in the medium would transform
the vector $\bfx = (100001000010000)$ to $(100001100010000)$
and the vector $\bfx = (000101011100010)$ to $\phi(\bfx) = (000001111100000)$.
Note that $\phi(\bfx) \neq \bfx$ iff a 01 or a 10 falls within some grain.
In particular, $\phi(\phi(\bfx)) = \phi(\bfx)$ for any $\x$.

In this paper, we consider only the case of granular media composed of grains
of length at most 2. Even this simplest possible case brings out the
complexity of the problem of coding to correct errors caused by this
combinatorial model. Most of the results we present can be extended
straightforwardly to the case of magnetic media with a more general grain
distribution.

Note that in a medium with grains of length at most 2, it is
precisely the length-2 grains that can cause bit errors.
We denote by $\Phi_{n,t}$ the set of operators $\phi$
corresponding to all such media with $n$ bit cells
and at most $t$ grains of length equal to 2. Then, for
$\bfx \in \{0,1\}^n$, we let
$\Phi_{n,t}(\bfx)=\{\phi(\bfx): \phi\in \Phi_{n,t}\}$,
and call two vectors $\bfx_1,\bfx_2\in \{0,1\}^n$ $t$-{\em confusable} if
   $$
    \Phi_{n,t}(\bfx_1)\cap \Phi_{n,t}(\bfx_2)\ne \emptyset.
   $$
A binary code $\cC$ of length $n$ is said to correct $t$ grain errors
if
no two distinct vectors $\bfx_1,\bfx_2\in \cC$ are $t$-confusable.

In Sections~II and III of this paper, we study properties of 
$t$-grain-correcting codes. We derive several bounds on the maximum size
of a length-$n$ binary code that corrects $t$ grain errors.
Our lower bounds are based on either explicit constructions or
existence arguments, while our upper bounds are based on 
the count of runs of identical symbols in a vector or 
on a clique partition of the ``confusability graph'' of the space $\{0,1\}^n.$ 
We also briefly consider list-decodable grain-correcting codes, 
and derive a lower bound on the maximum cardinality of such codes
by means of a probabilistic argument.

In Section~IV, we consider a scenario in which the locations of 
the grains are available to either the encoder or the decoder of the data, 
and derive estimates of the size of codes in this setting.

In Section~V, we consider a probabilistic channel model that corresponds
to the one-dimensional combinatorial model of errors discussed above, calling it
the ``grains channel''. We again confine ourselves to length-2 grains.
Our objective is to estimate the capacity of the channel.
For a lower bound on the capacity we restrict our attention 
to uniformly distributed, independent input letters which corresponds 
to the case of {\em symmetric information rate} (symmetric capacity or SIR) 
of the channel. We are able to find an exact expression for the SIR as 
an infinite series which gives a lower bound on the true capacity. 
To estimate capacity from above, we relate the grains channel to
an erasure channel in which erasures never occur in adjacent symbols,
and are otherwise independent. We explicitly compute the capacity of 
this erasure channel, and observe that the grains channel is a 
stochastically degraded version of the erasure channel. The capacity
of the erasure channel is thus an upper bound on the capacity 
of the grains channel.

We would like to acknowledge a concurrent independent paper by
Iyengar, Siegel, and Wolf \cite{ISW10} which contains some of our results
from Section~V. The authors of \cite{ISW10} considered
a more general channel model that includes our probabilistic model of the
grains channel as a particular case. Their paper contains results
that cover our Propositions~\ref{Cg1_prop} and \ref{zero_error_prop}, 
as well as our Theorem \ref{Cg_upbnd}. However, a major contribution of ours
that cannot be found in \cite{ISW10} is our Theorem~\ref{Cg_lobnd}, 
in which we give an exact expression for the SIR of the grains channel.

Throughout the paper, $h(x)=-x\log_2 x-(1-x)\log_2(1-x)$ denotes the binary
entropy function.

\section{Constructions of grain-correcting codes}\label{sect:constr}

As observed above, when the length of the grains does not exceed 2, 
bit errors are caused only by length-2 grains.
Furthermore, it can only be the second bit within such a grain that can be
in error. Thus, any code that can correct $t$ bit-flip errors (equivalently,
a code with minimum Hamming distance at least 2t+1) is a $t$-grain-correcting
code. In particular, $t$-grain-correcting codes whose parameters meet
the Gilbert-Varshamov bound (see e.g.\ \cite[p.\ 97]{R06})
are guaranteed to exist. But we can sometimes do better than
conventional error-correcting codes by taking advantage of the
special nature of grain errors.

Observe that the first bit to be written onto the medium can never be
in error in the grain model. So, we can construct $t$-grain-correcting
codes $\cC$ of length $n$ as follows: take a code $\cC'$ of length $n-1$
that can correct $t$ bit-flip errors, and set $\cC = (0|\cC') \cup (1|\cC')$.
Here, for $b \in \{0,1\}$, $(b|\cC')$ refers to the set of vectors
obtained by prefixing $b$ to each codevector of $\cC'$. For example,
when $n = 2^m$, we can take $\cC'$ to be the binary Hamming code
of length $2^m-1$, yielding a 1-grain-correcting code $\cC$
of size $|\cC| = 2^n/n$. Note that $2^n/n$ exceeds the sphere-packing 
(Hamming) upper bound, i.e., is greater than the cardinality
of the optimal binary single-error-correcting code
of length $n=2^m.$

More generally, again when $n$ is a power of 2, we can take $\cC'$
to be a binary BCH code of length $n-1$ that corrects $t$ bit-flip errors.
The above construction then yields a $t$-grain-correcting code $\cC$
of length $n$ and size $|\cC|\ge 2^n/n^t$.

We next describe a completely different, and remarkably simple, construction
of a length-$n$ grain-correcting code that corrects \emph{any} number
of grain errors. For even integers $n = 2m$, $m \geq 1$, define the code
$\cR_n \subset \{0,1\}^n$ as the set
  \begin{equation}\label{eq:Rn}
\{(x_1 x_2 \ldots x_{2m})\in\{0,1\}^n:
 x_{i-1} = x_i \text{ for all even indices } i\}.
  \end{equation}
Note that when a codevector from $\cR_n$ is written onto a medium
composed of grains of length at most 2, the bits at even coordinates
remain unchanged. Indeed, a bit at an even index $i$ could be in error
only if
a grain starts at index $i-1$, causing the bit at index $i-1$
to overwrite the bit at index $i$. However, the two bits are identical
by construction. Thus, $\cR_n$ is a code of size $2^{n/2}$
that corrects an arbitrary number of grain errors.
This construction can be extended to odd lengths $n = 2m+1$, $m \geq 1$,
as follows: $\cR_n = (0|\cR_{2m}) \cup (1|\cR_{2m})$.

\section{Bounds on the size of grain-correcting codes}\label{sect:bounds}

Let $M(n,t)$ denote the maximum size of a length-$n$ binary code
that is $t$-grain-correcting. The constructions of the previous section show
that $M(n,t) \geq 2^{\lceil n/2 \rceil}$ for any $n$ and $t$,
and $M(n,t) \geq 2^n/n^t$ when $n$ is a power of 2.
In an attempt to determine the tightness of these lower bounds,
we derive below some upper bounds on $M(n,t)$.

\subsection{Upper Bounds Based on Counts of Runs}\label{sect:runs}
Denote by $r(\bfx)$ the number of runs (maximal subvectors of 
consecutive identical symbols) in the vector $\bfx\in\{0,1\}^n.$
As remarked in Section~I, a single grain can change $\bfx$
to a different vector if and only if the grain straddles
the boundary between two successive runs in $\bfx$. Thus,
$|\Phi_{n,1}(\bfx)|=1+(r(\bfx)-1)=r(\bfx).$
For $t\ge 2$, the number $|\Phi_{n,t}(\bfx)|$ is not readily expressible in a
closed form. Nevertheless, we have the following lemma. \\[-20pt]
\begin{lemma} \label{phi_lobnd_lemma}
$$
|\Phi_{n,t}(\x)| \ge 1+\sum_{i=1}^t \frac{1}{i!}\prod_{j=0}^{i-1}(r(\x)-1-3j).
$$
\end{lemma}
\vspace*{-0.1in}
\begin{IEEEproof}
The right-hand side is a worst-case count of the number of ways
in which $i \leq t$ length-2 grains can be placed so that
each grain straddles the boundary between successive runs in $\bfx$.
The first grain can be placed in $r(\bfx)-1$ ways;
after that, in the worst case (which happens when the first grain falls
in the middle of a 1010 or 0101), the next grain can be placed in
$(r(\bfx)-1) - 3$ ways; and so on.
\remove{the first bit of the first one can
be chosen in $r(\bfx)-1$ ways, and in the worst case, the second grain can
be placed in $(r(\bfx)-1)-3$ ways so that its effect of $\bfx$ is nontrivial.
In general, the placement of the $i$th error, $i=1,\dots, t,$
can be chosen in at least $r(\bfx)-1-3(i-1)$ ways,
hence (\ref{lower_bound_on_phi_2}).}
\end{IEEEproof}

This leads to the following upper bound on $M(n,t)$. \\[-20pt]
\begin{theorem} \label{thm:const}
For any fixed value of $t$,
$$
M(n,t) \le \frac{2^n}{n^t} \, (t!\,2^t + 2 + o(1)),
$$
where $o(1)$ denotes a term that goes to 0 as $n \to \infty$.
\end{theorem}
\begin{IEEEproof}
Let $\cC$ be a $t$-grain-correcting code of length $n$,
and let 
  $$
   \cC_1 = \big\{\x \in \cC: |r(\x) - n/2| \le \sqrt{nt\log_2{n}}\big\}.
  $$
For any $\x \in \cC_1$, we have from Lemma~\ref{phi_lobnd_lemma},
\begin{eqnarray}
|\Phi_{n,t}(\x)| &\ge& \frac{1}{t!}(r(\x) -1 -3(t-1))^t \notag \\
&\ge& \frac{1}{t!}(n/2 - \sqrt{nt\log_2{n}} -1 -3(t-1))^t \label{ineq1} 
\end{eqnarray}
Since $\cC_1$ itself is $t$-grain-correcting, we also have
\begin{equation}
  2^n \geq \Big|\bigcup_{\x \in \cC_1} \Phi_{n,t}(\x)\Big|
 = \sum_{\x \in \cC_1} |\Phi_{n,t}(\x)|.
\label{ineq2}
\end{equation}
It follows from (\ref{ineq1}) and (\ref{ineq2}) that
  $$
   |\cC_1|\le \frac{2^{n+t} t!}{n^t} \,(1+o(1)).
  $$
Now, let $\cC_2=\cC\backslash\cC_1$.
We shall bound from above the size of
$\cC_2$ by the number of vectors $\x \in \{0,1\}^n$ such that
$|r(\x) - n/2| \ge \sqrt{nt\log_2{n}}.$ Define 
$\psi:\{0,1\}^n\to\{0,1\}^{n-1}$ by setting
  $$
     \psi((x_1,x_2,\dots,x_n))=(x_1 \oplus x_2,x_2 \oplus x_3,
      \ldots,x_{n-1}\oplus x_n)
  $$
where $\oplus$ denotes modulo-2 addition.
Then, $r(\bfx)=\wt(\psi(\bfx))+1$, where $\wt(\cdot)$ denotes Hamming weight.
For any given vector $\bfy\in\{0,1\}^{n-1}$, there are exactly two vectors
$\bfx_1,\bfx_2={\bf 1}\oplus\bfx_1$ such that $\psi(\bfx_1)=\psi(\bfx_2)=\bfy.$
Therefore,
  \begin{eqnarray*}
|\cC_2| &\le& 2|\{\y\in \mathbb{F}_2^{n-1} :|\wt(\y)+1 - n/2|
\ge \sqrt{nt\log_2{n}} \}|\\
&\le& 4 \sum_{i = 0}^{n/2-\sqrt{nt\log_2{n}}}{{n-1} \choose i}\\
&\le & 4 \exp\Big\{(n-1)h\Big(\frac{1}{2} -\frac{2\sqrt{nt\log_2{n}}-1}{2(n-1)}
\Big)\Big\},
\end{eqnarray*}
where $h(z)=-z\log_2z-(1-z)\log_2(1-z)$ is the binary entropy function.
Since $h(\frac{1}{2}-x) \le 1-\frac2{\ln 2}x^2$,
    \begin{eqnarray*}
|\cC_2| &\le&4
\exp\Big\{(n-1) -\frac2{\ln 2}\frac{(2\sqrt{nt\log_2{n}}-1)^2}{4(n-1)}\Big\}\\
&\le& {2^{n+1}}{n^{-t}}.
\end{eqnarray*}
We conclude by noting that $|\cC|=|\cC_1|+|\cC_2|.$
\end{IEEEproof}

For fixed $t$, the upper bound of the above theorem is within a constant
multiple of the lower bound $M(n,t) \ge 2^n/n^t$, stated earlier as
being valid when $n$ is a power of 2.

\remove{
Indeed, for $t=1$ we obtain
   $$
  M(n,1)\le\frac{2^{n+1}}n(1+o(1)).
   $$
At the same time, let $n=2^m$ and take $\cC=(0|\cH_m)\cup(1|\cH_m),$
where $\cH_m$ is the Hamming code of length $n-1,$ and where $(i|\cH_m)$
means a set of vectors obtained upon adding a single bit $i=0,1$ to each of
the Hamming codevectors. Then $|\cC|=2^n/n,$ and $|\cC|$ corrects any single
Hamming error in positions $2,3,\dots,n$ by using the Hamming code
correcting procedure. Therefore, $\cC$ is a single-grain-correcting code.

Generalizing, we can take $\cC=(0|\cB_m)\cup(1|\cB_m),$ where $\cB_m$
is a BCH code that corrects $t$ Hamming errors. By the same token, $\cC$
corrects $t$ grain errors, and $|\cC|\ge 2^n/n^t,$ where $n$ is a power of 2.
}

The bound of Theorem \ref{thm:const} is not useful when $t$ grows linearly
with $n$, say, $t= n\tau$ for $\tau \in (0,\half]$.
In this case, we define
     \begin{equation}
\oR(\tau) = \limsup_{n\to\infty} \frac{\log_2 M(n,\lfloor n \tau \rfloor)}{n}.
\label{Rbar_def}
    \end{equation}
An upper bound on $\oR(\tau)$ for small $\tau$ can be established by an argument
similar to the proof of the previous theorem.

 \begin{proposition}\label{prop:upper}
Let $x^* = x^*(\tau)$ be the smallest positive solution of
the following equation:
   $$
\h\Big(\frac{1-x}{2}\Big) + \frac{1-x}{4}\h\Big(\frac{4\tau}{1-x}\Big) =1.
$$
For $\tau \le 0.0706$, the following bound holds true:
  \begin{equation} \label{eq:lin-t}
  \oR(\tau)\le \h\Big(\frac{1-x^*}2\Big).
  \end{equation}
\end{proposition}

\vspace*{.02in}

\begin{IEEEproof}
The proof relies on a coarser estimate of $|\Phi_{n,t}(\x)|$ than the one in 
Lemma~\ref{phi_lobnd_lemma}. Consider the boundaries between the 
$(2i-1)$-th and $2i$-th runs in $\x$, $i = 1,2,\ldots,\lfloor r(\x)/2 \rfloor$.
Length-2 grains can be independently placed across these boundaries, 
leading to the lower bound
\begin{equation}
\label{lower_bound_on_phi_1}
|\Phi_{n,t}(\x)| \ge \sum_{i=0}^t{\lfloor r(\x)/2 \rfloor \choose i}.
\end{equation}

For $t = \lfloor\tau n\rfloor$, let $\cC$ be a $t$-grain-correcting code. 
For some $\delta > 0$, let
      $$
    \cC_1=\Big\{\x\in \cC: 
  	r(\x)/2 \ge \left\lfloor\frac{n}{4}(1-\delta)\right\rfloor\Big\}
      $$
The bound (\ref{lower_bound_on_phi_1}) implies that for each $\x \in \cC_1$
  $$
|\Phi_{n,t}(\x)|
 \ge \sum_{i=0}^t \binom{\lfloor \frac{n}{4}(1 -\delta) \rfloor}{i}.
  $$
From the above and (\ref{ineq2}), we obtain
  $$
|\cC_1| \le 
\frac{2^n}{\sum_{i=0}^{t} 
           \binom{ \lfloor\frac{n}{4}(1 -\delta)\rfloor} {i}}.
  $$
The size of the remaining subset of vectors $\cC_2=\cC\backslash \cC_1$
does not exceed the number of all vectors $\x$ with
$r(\x)\le\frac n2(1-\delta),$ i.e.,
  $$
   |\cC_2| \le \sum_{i = 0}^{\lfloor\frac{n}{2}(1 -\delta)\rfloor}\binom{n-1}i
 \le 2^{n\h\left(\frac{1-\delta}{2}\right)}.
  $$
Therefore,
   $$
  |\cC| \le \min_{\delta > 0}
\Big\{\frac{2^n}{\sum_{i=0}^{t}\binom{\lfloor\frac{n}{4}(1 -\delta)\rfloor}{i}}
+2^{n\h\left(\frac{1-\delta}{2}\right)} \Big\}.
   $$

When $\tau \le \frac{1-\delta}{8}$, or equivalently, $\delta \le 1-8\tau$,
the dominant term in the sum in the denominator above is 
$\binom{\lfloor\frac{n}{4}(1 -\delta)\rfloor}{\lfloor\tau n\rfloor}$,
which is bounded below by $\frac{1}{\sqrt{8n}}
2^{\frac{n(1-\delta)}{4}\h\left(\frac{4\tau}{1-\delta}\right)}$. 
From this, we obtain
\begin{equation}
  \oR(\tau)\le\min_{0 < \delta \leq 1 - 8 \tau}\max\Big\{
1-\frac{1-\delta}4 \h\Big(\frac {4\tau}{1-\delta}\Big),
\h\Big(\frac{1-\delta}2\Big)\Big\}
\label{eq:oR}
\end{equation}

Now, for $1-8\tau$ to be positive, we need $\tau < 1/8$. 
For any fixed $\tau \in [0,1/8)$, and $\delta \in [0,1-8\tau]$, the function 
$f(\delta) = 1-\frac{1-\delta}4 \h\Big(\frac {4\tau}{1-\delta}\Big)$
is an increasing function of $\delta$, while the function 
$g(\delta) = \h(\frac{1-\delta}{2})$ is a decreasing function of $\delta$.
At $\delta = 0$, we have $g(\delta) \ge f(\delta)$. If, at $\delta = 1-8\tau$,
we have $g(\delta) \le f(\delta)$, then it follows that the minimum over 
$\delta$ in (\ref{eq:oR}) is achieved when $f(\delta) = g(\delta)$. 
In other words, the minimizing value of $\delta$ in this case 
is precisely the $x^*$ in the statement of the proposition. It is readily
verified that at $\delta = 1-8\tau$, we have 
$g(\delta) - f(\delta) = h(4\tau) + 2\tau - 1$, 
which is negative when $\tau \leq 0.0706$. 
\hfill\end{IEEEproof}

\medskip

\begin{figure}
\centering
\includegraphics[width=0.5 \textwidth]{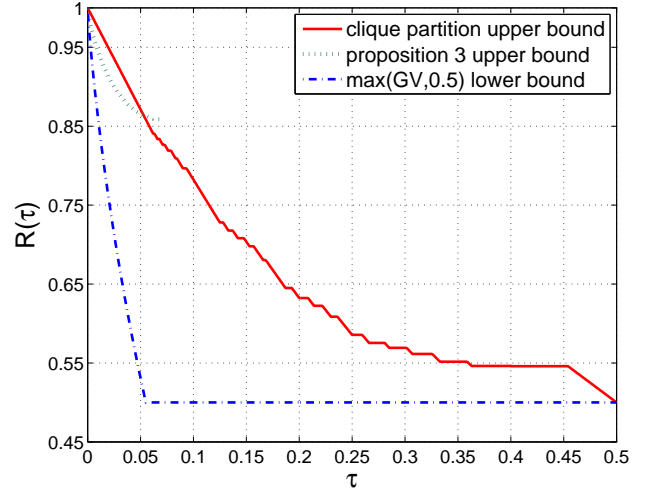}
\vspace*{-12pt}
\caption{Upper and lower bounds on the asymptotic coding rate
of grain-correcting codes.}\label{fig1}
\end{figure}

Bound (\ref{eq:lin-t}) is plotted in Fig.~\ref{fig1},
along with the asymptotic version of the Gilbert-Varshamov lower bound, which, 
as observed in Section~\ref{sect:constr},
is also valid for grain-correcting codes. The methods of the next subsection
yield upper bounds on $\oR(\tau)$ for any $\tau \leq \half$, but these are
harder to evaluate than the bound of Proposition~\ref{prop:upper}.

\subsection{Upper Bounds Based on Clique Partitions}\label{sect:cliques}
A \emph{clique partition} of a graph $G$ is a partition
$(V_1,\ldots,V_k)$ of its vertex set $V$ such that the subgraph
induced by each $V_j$, $j = 1,\ldots,k$, is a clique of $G$.
Let $\bar{\chi}(G)$ denote the smallest size (number of parts)
of any clique partition of $G$.

Let $G(n,t)$ be a confusability graph of the code space, defined as follows:
the vertex set of $G(n,t)$ is $\{0,1\}^n$, and two distinct vertices
$\x,\x'$ are joined by an edge iff they are $t$-confusable.
\remove{Clearly,
$M(n,t)$ is equal to the independence number
({i.e.}, the maximum size of an independent set) of $G(n,t)$.
However, we need $G(n,t)$ for a different purpose.}
For notational simplicity, we denote $\bar{\chi}(G(n,t))$ by
$\bar{\chi}_{n,t}$. We do not assume that $t$ is an integer;
for non-integer values of $t$, we set
$\bar{\chi}_{n,t} = \bar{\chi}_{n,\lfloor{t}\rfloor}$.


To state our next result, we need to extend the definition of $M(n,t)$
as follows: $M(0,t) = 1$ for all $t$. \\[-20pt]

\begin{proposition}
For $m \leq n$ and $s \leq t$,
$$
M(n,t) \leq \bar{\chi}_{m,s} \, M(n-m,t-s).
$$
\label{basic_prop}
\end{proposition}
\vspace*{-.3in}
\begin{IEEEproof}
Let $\cC \subseteq \{0,1\}^n$ be a $t$-grain-correcting code of size
$|\cC| = M(n,t)$, and let $(V_1,\ldots,V_k)$ be a clique partition
of $G(m,s)$ of size $k = \bar{\chi}_{m,s}$. For $j = 1,\ldots,k$,
define $\cC_j = \{(c_1,\ldots,c_n) \in \cC: (c_1,\ldots,c_m) \in V_j\}$.
As the $V_j$'s form a partition of $\{0,1\}^m$, the $\cC_j$'s form
a partition of $\cC$. Therefore, it is enough to show that
$|\cC_j| \leq M(n-m,t-s)$ for all $j$.
\remove{for we would then have
$|\cC| = \sum_{j=1}^k |\cC_j| \leq k M(n-m,t-s)$, which would prove the
proposition.}
Let $\cC'_j = \{(c_{m+1},\ldots,c_n):
  \exists \, (c_1, \ldots, c_m, c_{m+1}, \ldots, c_n) \in \cC_j\}$.
The canonical projection map $\pi: \cC_j \to \cC'_j$ is a bijection;
to see this, it is enough to show that $\pi$ is injective.
If $\pi(\c) = \pi(\hbc)$
for $\c,\hbc \in \cC_j$, then $\c = (c_1,\ldots,c_m,c_{m+1},\ldots,c_n)$
and $\hbc = (\hc_1,\ldots,\hc_m,c_{m+1},\ldots,c_n)$ for some
$(c_1,\ldots,c_m)$ and $(\hc_1,\ldots,\hc_m)$ in $V_j$.
But, since the subgraph induced by $V_j$ forms a clique in $G(m,s)$,
we have that $(c_1,\ldots,c_m)$ and $(\hc_1,\ldots,\hc_m)$ are $s$-confusable.
Thus, we see that $\c,\hbc$ are $s$-confusable
(and hence $t$-confusable since $s \leq t$) unless $\c = \hbc$.
Hence, $\pi$ is a bijection, so that $|\cC_j| = |\cC'_j|$.

We further claim that $\cC'_j \subseteq \{0,1\}^{n-m}$ is a
$(t-s)$-grain-correcting code, which would show that
$|\cC_j|=|\cC'_j| \leq M(n-m,t-s)$.
 Indeed, consider any pair of distinct words
$\c',\bfd' \in \cC'_j$. There exist distinct codewords
$(\bfa',\c')$ and $(\bfb',\bfd')$ in $\cC_j$.
By definition of $\cC_j$, $\bfa'$ and $\bfb'$ are $s$-confusable.
So, if $\c'$ and $\bfd'$ were $(t-s)$-confusable, then
$(\bfa',\c')$ and $(\bfb',\bfd')$ would be $t$-confusable,
which cannot happen for distinct codewords in $\cC_j$.
Hence, $\cC'_j$ is a $(t-s)$-grain-correcting code.
\end{IEEEproof}

\medskip

If $n/m \geq t/s$ (or equivalently, $t/n \leq s/m$),
then repeated application of the above proposition yields
$$M(n,t) \leq {(\bar{\chi}_{m,s})}^{\lfloor t/s \rfloor}
                M(n-m{\lfloor t/s \rfloor},t-s{\lfloor t/s \rfloor}),$$
from which we obtain the following corollary. \\[-16pt]

\begin{corollary}
If $t/n \leq s/m$, then
$$
M(n,t) \leq
    {(\bar{\chi}_{m,s})}^{\lfloor t/s \rfloor} \, 2^{n-m{\lfloor t/s \rfloor}}.
$$
\label{cor1}
\end{corollary}

\vspace*{-.2in} It is difficult to determine
$\bar{\chi}_{m,s}$ exactly for arbitrary $m,s$. Upper bounds
on $\bar{\chi}_{m,s}$ can be found by explicit constructions
of clique partitions of $G(m,s)$. Observe that for any $\y \in \{0,1\}^m$,
the set $\Phi_{m,s}^{-1}(\y) := \{\x \in \{0,1\}^m: \y \in \Phi_{m,s}(\x)\}$
forms a clique in $G_{m,s}$. Thus, clique partitions of size $k$
can be found by identifying sequences $\y_1,\ldots,\y_k \in \{0,1\}^m$
such that the sets $\Phi_{m,s}^{-1}(\y_j)$, $j = 1,\ldots,k$,
cover $\{0,1\}^m$. Note that the sets
$V_j = \Phi^{-1}_{m,s}(\y_j) \setminus \left(\bigcup_{i < j} V_i\right)$,
$j = 1,\ldots,k$, then form a clique partition of $G(m,s)$.
We implemented the greedy algorithm described below
to find such a list of sequences $\y_1,\ldots,\y_k$,
and hence, a clique partition $V_1,\ldots,V_k$.

\begin{algorithm}[ht]
\caption{A greedy algorithm for finding clique partitions in $G(m,s)$.}
\begin{algorithmic}[1]
\STATE 
determine the sets $\Phi^{-1}_{m,s}(\y)$
for all $\y \in \{0,1\}^m$;
\STATE set $B(\y) = \Phi^{-1}_{m,s}(\y)$ for all $\y \in \{0,1\}^m$, \\
set $k = 0$;
\STATE \algorithmicwhile \; there exists a $\y$ such that $B(\y)$ is non-empty
\algorithmicdo
\STATE \ \ \ \ \ \ \ \ $k \leftarrow k+1$ ;
\STATE \ \ \ \ \ \ \ \ find a $\y_k$ such that
$|B(\y_k)| = \max_{\y \in \{0,1\}^m} |B(\y)|$;
\STATE \ \ \ \ \ \ \ \ set $V_k = B(\y_k)$;
\STATE \ \ \ \ \ \ \ \ \algorithmicfor\; each $\y \in \{0,1\}^m$
\STATE \ \ \ \ \ \ \ \ \ \ \ \ \ \ \ \
$B(\y) \leftarrow B(\y) \setminus V_k$;
\STATE \algorithmicreturn\; $V_1,\ldots,V_k$.
\end{algorithmic}
\end{algorithm}

Table~\ref{table:chi_bounds} lists upper bounds on $\bar{\chi}_{m,s}$
obtained via our implementation of the greedy algorithm. The underlined
entries in the table are known to be exact values of $\bar{\chi}_{m,s}$,
obtained either from the fact that
$\bar{\chi}_{m,s} \ge M(m,s) \ge 2^{\lceil m/2 \rceil}$,
or from specialized arguments that we omit here.

\begin{table*}[!t]
\begin{center}
\begin{tabular}{cc||c|c|c|c|c|c|c|c|c|c|c|c|c|c|c|}
& & \multicolumn{15}{c|}{$m$} \\ \cline{3-17}
& & 2 & 3 & 4 & 5 & 6 & 7 & 8 & 9 & 10 & 11 & 12 & 13 & 14 & 15 & 16 \\
\hline\hline
\multicolumn{1}{c|}{\multirow{3}{*}{$s$}} &
\multicolumn{1}{|c||}{1} & \underline{2} & \underline{4} & \underline{6} & \underline{10} & 18 & 36 & 66 & 122 & 236 & 428 & 834 & 1574 & 3008 & 5716 & 11014 \\
\cline{2-17}
& \multicolumn{1}{|c||}{2} &  &  & \underline{4} & \underline{8} & 12 & 18 & 30 & 54 & 92 & 162 & 284 & 530 & 948 & 1730 & 3210 \\
\cline{2-17}
& \multicolumn{1}{|c||}{3} &  &  &  &  & \underline{8} & \underline{16} & 24 & 34 & 56 & 88 & 138 & 238 & 418  & 716 & 1266 \\
\cline{2-17}
& \multicolumn{1}{|c||}{4} &  &  &  &  &  &  & \underline{16} & \underline{32}& 44 & 64 & 98 & 156 & 248 & 392 & 662  \\
\cline{1-17}
\end{tabular}\end{center}
\caption{Upper bounds on $\bar{\chi}_{m,s}$ obtained by computer search;
the underlined table entries are known to be exact values of
$\bar{\chi}_{m,s}$.}
\label{table:chi_bounds}
\end{table*}

From Corollary~\ref{cor1} and Table~\ref{table:chi_bounds}, we can obtain
a suite of upper bounds on $M(n,t)$ valid for various ranges of $n$ and $t$;
for example, the entry for $(m,s) = (10,1)$ in the table yields that
$M(n,t) \leq 236^t 2^{n-10t}$ for $t/n \leq 1/10$. The following upper bound
on $\oR(\tau)$, which was defined in (\ref{Rbar_def}), is also a direct
consequence of Corollary~\ref{cor1}.

\begin{corollary}
For $m,s$ such that $\tau \leq s/m$,
$$
\oR(\tau)
 \leq 1 - \tau\left(\frac{m}{s} - \frac{1}{s} \log_2 \bar{\chi}_{m,s} \right).
$$
\label{cor2}
\end{corollary}
\vspace*{-0.2in}

When used in conjunction with Table~\ref{table:chi_bounds}, the above corollary
gives useful upper bounds on $\oR(\tau)$. For instance, using the table entry
for $(m,s) = (16,4)$, we find that
 $\oR(\tau) \leq 1 - \tau(4 - \frac{1}{4} \log_2 662) \approx 1 - 1.657 \tau$
for $\tau \leq 1/4$. Figure~\ref{fig1} plots the minimum of all the
upper bounds on $R(\tau)$ obtainable from Corollary~\ref{cor2}
and the entries of Table~\ref{table:chi_bounds}.

Setting $s = \tau m$ in Corollary~\ref{cor2}, we obtain
$\oR(\tau) \leq \frac{1}{m} \log_2 \bar{\chi}_{m,\tau m}$, and hence,
\begin{equation}
\oR(\tau) \leq \inf_m \frac{1}{m} \log_2 \bar{\chi}_{m,\tau m}
= \lim_{m\to\infty} \frac{1}{m} \log_2 \bar{\chi}_{m,\tau m}.
\label{Rbar_upbound}
\end{equation}
The last equality above follows from Fekete's lemma
(see e.g.\ \cite[p.~85]{lin92}), noting that $f(m) = \log_2\bar\chi_{m,rm}$
is a subadditive function, i.e., $f(m+n)\le f(m)+f(n).$
The bound in (\ref{Rbar_upbound}) is presently only of theoretical interest,
as the infimum (or limit) on the right-hand side is difficult
to evaluate in general.

\subsection{A List-Decoding Lower Bound\label{sec:listdec}}

We briefly venture into the territory of list-decoding in this section,
and give a lower bound on the achievable coding rate of a list-$L$-decodable
code. Recall that in the list-decoding setting, the decoder is allowed
to produce a list of up to $L$ codewords. Formally, a code $\cC$
is \emph{list-$L$ $t$-grain-correcting} if for any vector
$\x \in \{0,1\}^n$, $|\{\c \in \cC: \x \in \Phi_{n,t}(\c)\}| \le L.$
In words, for any received vector $\x \in \{0,1\}^n$, there are at
most $L$ codewords that could get transformed to $\x$ by the action
of an operator $\phi \in \Phi_{n,t}$.

We will find the following definition useful in what is to follow.
For $\phi \in \Phi_{n,t}$, let $\bfe_\phi$ be the vector
$(e_1,\ldots,e_n) \in \{0,1\}^n$, with $e_j = 1$ iff $\phi$
has a length-2 grain beginning at the $(j-1)$th bit cell.
Define $\cE_{n,t} = \{\e_{\phi}: \phi \in \Phi_{n,t}\}$.
Note that $\cE_{n,t}$ consists of all binary ``error vectors''
of length $n$ and Hamming weight at most $t$ such that the
first coordinate is always $0$ and no two $1$'s are adjacent.
An easy counting argument shows that
\begin{equation}
|\cE_{n,t}| = \sum_{i=0}^t{ {n-i}  \choose i}.
\label{eq:Ent}
\end{equation}

Denote by $M(n,t;L)$ the maximum size of a list-$L$ $t$-grain-correcting code
of length $n$, and define for $0 \leq \tau \leq \half$,
$$
\uR(\tau;L) 
 = \liminf_{n\to \infty}\frac{\log_2{M(n,\lfloor n\tau \rfloor;L)}}{n}.
$$

\begin{proposition}
We have
$$
M(n,t;L) \ge \frac{2^{{nL}/(L+1)}}{\sum_{i=0}^t\binom{n-i}{i}},
$$
and hence,
$$
\uR(\tau;L) 
 \ge \frac L{L+1} - (1-\tau)\h\Big(\frac{\tau}{1-\tau}\Big)
$$
for $\tau \le \frac12 - \frac{\sqrt{5}}{10} \approx 0.2764.$
\label{listdec_prop}
\end{proposition}
\begin{IEEEproof}
For a vector $\x \in \{0,1\}^n$ let us define
$$
B(\x)= \{\z \in \{0,1\}^n: \x \in \Phi_{n,t}(\z)\}.
$$
Note that $B(\x) \subseteq \{\x \oplus \e: \e \in \cE_{n,t}\}$,
so that $|B(\x)| \leq |\cE_{n,t}| = \sum_{i=1}^n \binom{n-i}{i}$.

Let us construct the code by choosing 
$M$ codewords randomly and uniformly with replacement 
from $\{0,1\}^n.$
For a fixed vector $\y \in  \{0,1\}^n,$ call the choice of any $L+1$
codewords $\c_1,\dots,\c_{L+1}$ `bad' if 
$\c_1,\dots,\c_{L+1} \in B(\y)$. 
Clearly, the expected number of bad choices for a random code $\cC$
is less than or equal to
\begin{align*}
2^n \binom M{L+1}
  &\bigg(\frac{\sum_{i=0}^t\binom{n-i}{i}}{2^n}\bigg)^{L+1}\\&<
\Big(M\sum_{i=0}^t\binom{n-i}{i}\Big)^{L+1} 2^{-nL} .
\end{align*}
Take $M={2^{{nL}/(L+1)}}/{\sum_{i=0}^t\binom{n-i}{i}},$
then the ensemble-average number of bad $(L+1)$-tuples is less than 1.
Therefore there exists a code of size $M$ in which all the $(L+1)$-tuples
of codewords are good.
This implies the lower bound on $M(n,t;L)$.

The bound on $\uR(\tau;L)$ follows from the observation that
$\binom{n-i}{i}$ increases with $i$ for 
$i\le \frac{1}{10}(5n+3-\sqrt{5n^2+10n+9}).$
Thus, as long as $t/n \leq \frac{1}{2} - \frac{\sqrt{5}}{10},$
the asymptotics of the summation $\sum_{i=0}^t \binom{n-i}{i}$
is determined by the term $\binom{n-t}{t}$.
\end{IEEEproof}

\medskip

We do not at present have a useful upper bound on $M(n,t;L)$.

\section{Grain pattern known to encoder/decoder}
In this section, we assume that the user of the recording system
is capable of testing the medium and acquiring information about the
structure of its grains. This information is used for the writing of
the data on the medium or performing the decoding. Specifically, we assume
again a medium with $n$ bit cells and at most $t$ grains of length 2,
but now the locations of the grains are available either to the decoder
but not the encoder of the data (Scenario I) or, conversely, to the
encoder but not the decoder (Scenario II). Accordingly,
let $M_i(n,t), i=1,2$, be the maximum number of messages
that can be encoded and decoded without error
in each of the two scenarios. Also, for $0\le\tau\le\half$, let
  $$
    \uR_i(\tau)=\liminf_{n\to\infty}
      \frac{\log_2 M_i(n,\lfloor n\tau \rfloor)}{n}, \quad i=1,2,
  $$
be the coding rate achievable in each situation when $t$ grows proportionally
with $n$, with constant of proportionality $\tau$.

For the analysis to follow, we need to recall the definition of $\cE_{n,t}$
from Section~\ref{sec:listdec}, and the fact (\ref{eq:Ent}) that 
$|\cE_{n,t}| = \sum_{i=0}^t \binom{n-i}{i}$.

\subsection{Scenario I}

Here, we assume that the locations of the grains are known to the decoder
of the data but are not available at the time of writing on the medium.
A code $\cC$ is said to correct $t$ grains known to the receiver
if $\phi(\bfx_1)\ne \phi(\bfx_2)$ for any two distinct vectors
$\bfx_1,\bfx_2\in \cC$ and any $\phi\in \Phi_{n,t}$.

An obvious solution for the decoder is to consider as erasures
the positions that could be in error, so the encoder can rely on a
$t$-erasure-correcting code. Therefore, by the argument of the
Gilbert-Varshamov bound,
$
M_1(n,t) \ge \frac{2^n}{\sum_{i=0}^t{ n \choose i}},
$
and hence,
$\uR_1(\tau) \ge 1-h(\tau)$.
However, this lower bound can be improved, as our next proposition shows.

\begin{proposition} \label{prop:known} We have
$$
M_1(n,t) \ge \frac{2^n}{\sum_{i=0}^t\binom{n -i} i}.
$$
Hence,
$\uR_1(\tau) \ge 1-(1-\tau)h(\frac{\tau}{1-\tau})$
for $\tau \le \frac{1}{2} - \frac{\sqrt{5}}{10} \approx 0.2764$.
\end{proposition}
\vspace*{-0.02in}
\begin{IEEEproof}
We shall construct a code $\cC$ of size at least $2^n/|\cE_{n,t}|$
by a greedy procedure. We begin with
an empty set, choose an arbitrary vector $\x_1$ and include it in $\cC$.
Having picked $\x_1,\ldots,\x_{i-1}$, for some $i \geq 1$, we choose
$\x_i$ so that
$$
\x_i \notin \bigcup_{j=1}^{i-1}\{\x_j \oplus \e: \e \in \cE_{n,t}\}.
$$
We stop when such a choice is not possible. At that point, we will have
constructed a code $\cC$ that satisfies $|\cC| \cdot |\cE_{n,t}| \ge 2^n.$

\remove{
The set $\cE_{n,t}$ consists of all binary vectors of length $n$ and Hamming weight
$\le t$ such that the first position is always $0$ and no two $1'$s
are adjacent. The number of such vectors is equal to the number of
partitions of the number $n$ into a sum of $1$s and $2$s with no more than
$t$  $2$s. It is easy see that $|\cE_{n,t}| = \sum_{i=0}^t{ {n -i}  \choose i}.$
}

We claim that $\cC$ corrects $t$ grains known to the receiver. Suppose not;
then there exists a grain pattern $\phi\in \Phi_{n,t}$ such that
$\phi(\bfx_i)=\phi(\bfx_j)$ for some $\bfx_i,\bfx_j \in \cC$, $i>j.$
Equivalently, $\bfx_i \oplus \e = \bfx_j \oplus \e'$ for some
error vectors $\e,\e'$ with $\text{supp}(\e),\text{supp}(\e')
\subseteq \text{supp}(\e_\phi)$, where $\text{supp}(\cdot)$
denotes the support of a vector.
We then have $\bfx_i = \bfx_j \oplus (\e \oplus \e')$ with
$\e \oplus \e' \in \cE_{n,t}$, which contradicts the construction of $\cC$.

As in the proof of Proposition~\ref{listdec_prop}, 
the bound on $\uR_1(\tau)$ follows from the observation that
when $t/n \leq \frac{1}{2} - \frac{\sqrt{5}}{10},$
the asymptotics of the summation $\sum_{i=0}^t \binom{n-i}{i}$
is determined by the term $\binom{n-t}{t}$.
\end{IEEEproof}

\subsection{Scenario II}

This scenario is  similar in spirit to the channel with {\em localized
errors} of Bassalygo et al.~\cite{BGP89}. In that setting, 
both the transmitter and the receiver know that all but $t$ positions of the
codevector will remain error-free, and the coordinates of the
$t$ positions which can (but need not) be in error are known to the
transmitter but not the receiver. Thus, in our Scenario~II, the
encoder may rely on codes that correct localized errors,
which according to \cite{BGP89} gives the bound
$\uR_2(\tau)\ge 1-h(\tau).$ Again, this bound can be improved. \\[-18pt]

\begin{proposition}
\label{lower_bound_scenario2}
We have
\begin{equation*}
M_2(n,t)\ge\frac{1}{2n}\frac{2^n}{\sum_{i=0}^t{ {n -i}  \choose i}}.
\end{equation*}
Hence,
$\uR_2(\tau) \ge 1-(1-\tau)h(\frac{\tau}{1-\tau})$
for $\tau \leq \frac{1}{2} - \frac{\sqrt{5}}{10} \approx 0.2764$.
\end{proposition}
\vspace*{-0.08in}
\begin{IEEEproof}
We show that when the encoder knows the error locations, then
it can successfully transmit
\begin{equation}
\label{lower_bound_nonas_2}
 M \ge\frac{1}{2n}\frac{2^n}{|\cE_{n,t}|}
 \end{equation}
messages to the decoder, which proves the claimed lower bound on $M_2(n,t)$.
We follow the proof of Theorem~$3$ of \cite{BGP89}.

Given a message $i \in \{1,\ldots,M\}$ to be transmitted, the transmitter will
use knowledge of the grain pattern $\phi$ (with $\e_\phi \in \cE_{n,t}$)
to encode $i$ using a suitably chosen vector from a set of binary vectors
$\cX^i = \{x_j^i: j = 1,\ldots,n\}$.
A vector $\x^i_j$ is said to be {\em good} for $\e \in \cE_{n,t}$
if for any $i \ne i'$ and for any $j'$ we have,
$$
\dist(\x^i_j \oplus \e,\x^i_j) < \dist(\x^i_j \oplus \e, \x^{i'}_{j'} ),
$$
where $\dist(\cdot,\cdot)$ denotes Hamming distance. The family of sets
$\cX^i$, $i = 1,\ldots,M$,
is {\em good}  if
for any $i \in \{1,\ldots,M\}$ and for any $\e \in \cE_{n,t}$,
there exists a vector $\x^i_j \in \cX^i$  that is good for $\e.$
A good family of sets $\cX^i$, $i = 1,\ldots,M$, enables the encoder to
transmit any message in $\{1,\ldots,M\}$ with perfect recovery by
the decoder. Indeed, given the grain pattern $\phi$,
the encoder chooses for transmission of message $i$ a vector
in $\cX^i$ that is good for $\bfe_\phi$.

Thus, we only need to show that for $M$ satisfying (\ref{lower_bound_nonas_2})
there exists a good family of sets $\cX^i = \{x^i_j: j=1,\ldots,n\}$,
$i = 1,\ldots,M$. There are $2^{n^2M}$ families of $M$ sets $\cX^i$,
each containing at most $n$ binary vectors of length $n$. Of these,
the number of families that are \emph{not} good does not exceed
$$
M\cdot|\cE_{n,t}|\cdot\left((M-1)n |\cE_{n,t}|\right)^n\cdot 2^{n^2(M-1)}.
$$
If $M$ satisfies (\ref{lower_bound_nonas_2}) with equality, then this
number is less than $2^{n^2M}$. Therefore, there exists a
good family of sets $\cX^i$.

The argument for the lower bound on $\uR_2(\tau)$ is the same as that
given for $\uR_1(\tau)$ in the proof of Proposition~\ref{prop:known}, since the
extra multiplicative factor of $\frac{1}{2n}$ does not affect the asymptotic
behavior. \end{IEEEproof}

\remove{\begin{figure}
\centering
\includegraphics[width=.5\textwidth]{known_grain_rate_bounds.eps}
\vspace*{-12pt}
\caption{Upper and lower bounds on the asymptotic coding rate 
in Scenarios~I and II.}\label{fig2}
\end{figure}

\medskip}

To summarize, we obtain a lower bound on $\uR_i(\tau)$, $i = 1$ or $2$,
of the form
$$
\uR_i(\tau)\ge\max\Big\{0.5,\, 1-(1-\tau)h(\frac{\tau}{1-\tau})\Big\}.
$$
This is because the rate-$\half$ code $\cR_n$ defined in (\ref{eq:Rn}) 
is still viable in the context of Scenarios~I and II.
A straightforward upper bound $\uR_i(\tau)\le 1-\tau$
follows from the fact that $M_1(n,t)$ and $M_2(n,t)$ 
cannot exceed $2^{n-t}$, which is simply the one-bit-per-grain upper bound. 

\section{Capacity of the Grains Channel}
Thus far in this paper, we have considered a combinatorial model of the one-dimensional
granular medium, and given various bounds on the rate of
$t$-grain-correcting codes. We will now switch to a parallel track by defining
a natural probabilistic model of a channel corresponding to the one-dimensional
granular medium with grains of length at most 2 (the ``grains channel''). 
This is a binary-output channel that can make an 
error only at positions where a length-2 grain ends. 
In fact, error events are data-dependent:
an error occurs at a position where a length-2 grain ends if and only if
the channel input at that position differs from the previous channel input.
Our goal is to estimate the Shannon-theoretic capacity for the grains channel model.
 Let us proceed to formal definitions.

Suppose $\bfx = x_1 x_2.. \ldots $ and  $\bfy = y_1 y_2.. \ldots $
 denote the input and output sequence respectively, with $x_i, y_i \in \{0,1\}$
for all $i.$ We further define the
sequence $\bfu = u_1  u_2.. \ldots$, where $u_i = 1$
(resp.\ $u_i = 0$) indicates that a length-$2$ grain
ends (resp.\ does not end) at position $i$. We take $\bfu$ to be
a first-order Markov chain, independent of the channel input $\bfx$,
having transition probabilities
$P(u_i | u_{i-1})$ as tabulated below (for some $p \in [0,1]$):
\begin{equation}
\text{
\begin{tabular}{c|cc}
 & $u_i = 0$ & $u_i = 1$ \\ \hline
$u_{i-1} = 0$ & $1-p$ & $p$ \\
$u_{i-1} = 1$ & $1$ & $0$
\end{tabular}
}.
\label{transit_probs}
\end{equation}
The grains channel makes an error at position $i$ (\emph{i.e.}, $x_i \neq y_i$)
if and only if $u_i = 1$ and $x_i \ne x_{i-1}$. To be precise,
\begin{equation}
y_i = x_i \oplus (x_i\oplus x_{i-1}) u_i,
\label{xi_yi_eq1}
\end{equation}
where the operations are being performed modulo 2.
Equivalently,
\begin{equation}
y_i =
\begin{cases} x_i & \text{ if } u_i = 0 \\
x_{i-1} & \text{ if } u_i = 1.
\end{cases}
\label{xi_yi_eq2}
\end{equation}
We will find it useful to define the error sequence $\z = z_1,z_2,z_3,\ldots$,
where $z_i = x_i \oplus y_i.$ Thus,
\begin{equation}
z_i = u_i (x_i \oplus x_{i-1}).
\label{zi}
\end{equation}
The case $i=1$ is not covered by the above definitions. We will include it
once we define a finite-state model of the grains channel.

The grains channel as we have defined above
is a special case of a somewhat more general ``write channel'' model considered
in \cite{ISW10}.

\subsection{Discrete Finite-State Channels\label{sec:dfsc}}

For easy reference, we record here some important facts about
discrete finite-state channels. The material in this section is
substantially based upon \cite[Section~4.6]{Gallager1968}.

A stationary \emph{discrete finite-state channel (DFSC)} has an input sequence
$\bfx = x_1,x_2,x_3,\ldots$, an output sequence $\bfy = y_1,y_2,y_3,\ldots$,
and a state sequence $\bfs = s_1,s_2,s_3,\ldots$. Each $x_n$ is a symbol
from a finite input alphabet $\cX$, each $y_n$ is a symbol
from a finite output alphabet $\cY$, and each state $s_n$
takes values in a finite set of states $\cS$. The channel is
described statistically by specifying a conditional probability assignment
$P(y_n,s_n | x_n,s_{n-1})$, which is independent of $n$. It is assumed that,
conditional on $x_n$ and $s_{n-1}$, the pair $y_n,s_n$ is statistically
independent of all inputs $x_j$, $j < n$, outputs $y_j$, $j < n$,
and states $s_j$, $j < n-1$. To complete the description of the channel,
an initial state $s_0$, also taking values in $\cS$, must be
specified.

For a DFSC, we define the \emph{lower} (or \emph{pessimistic}) \emph{capacity}
$\underline{C} = \lim_{n\to\infty} \underline{C}_n$,
and \emph{upper} (or \emph{optimistic}) \emph{capacity}
$\overline{C} = \lim_{n\to\infty} \overline{C}_n$, where
     \begin{align*}
          \underline{C}_n &= n^{-1} \max_{Q^n(\bfx^n)} \min_{s_0\in \cS}
I(\bfx^n;\bfy^n \mid s_0) \notag \\ 
\overline{C}_n &= n^{-1} \max_{Q^n(\bfx^n)} \max_{s_0\in \cS}
I(\bfx^n;\bfy^n \mid s_0). \notag 
      \end{align*}
In the above expressions, $I(\bfx^n;\bfy^n \mid s_0)$
is the mutual information between the length-$n$ input
$\bfx^n = (x_1,\ldots,x_n)$ and the length-$n$ output
$\bfy^n = (y_1,\ldots,y_n)$,
given the value of the initial state $s_0,$ and the maximum is taken
over probability distributions $Q^n(\bfx^n)$ on the input $\bfx^n$.
The limits in the above definitions of $\underline{C}$ and $\overline{C}$
are known to exist. Clearly, $\underline{C}_n \le \overline{C}_n$ for all $n$,
and thus, $\underline{C} \le \overline{C}$. The capacities
$\underline{C}$ and $\overline{C}$ have an operational meaning
in the usual Shannon-theoretic sense --- see Theorems~4.6.2 and 5.9.2
in \cite{Gallager1968}.

The upper and lower capacities coincide for a large class of channels
known as \emph{indecomposable} channels. Roughly, an indecomposable DFSC
is a DFSC in which the effect of the initial state $s_0$ dies away with time.
Formally, let $q(s_n \mid \bfx^n,s_0)$ denote the conditional probability
that the $n$th state is $s_n$, given the input sequence
$\bfx^n = (x_1,\ldots,x_n)$ and initial state $s_0$. Evidently,
$q(s_n \mid \bfx^n,s_0)$ is computable from the channel statistics.
A DFSC is indecomposable if, for any $\eps > 0$, there exists an $n_0$
such that for all $n \ge n_0$, we have
$$
|q(s_n \mid \bfx^n,s_0) - q(s_n \mid \bfx^n, s_0')| \le \eps
$$
for all $s_n$, $\bfx^n$, $s_0$ and $s_0'$. Theorem~4.6.3 of \cite{Gallager1968}
gives an easy-to-check necessary and sufficient condition for a DFSC to be
indecomposable: for some fixed $n$ and each $\bfx^n$, there exists a choice
for $s_n$ (which may depend on $\bfx^n$) such that
\begin{equation}
\min_{s_0} q(s_n \mid \bfx^n,s_0) > 0. 
\label{indecomp_cond}
\end{equation}
We note here that the channels we consider in the subsequent sections
are indecomposable except in very special cases. For these special cases,
it can still be shown that $\underline{C} = \overline{C}$ holds.

We make a few comments about DFSCs for which $\underline{C} = \overline{C}$
holds. We denote by $C$ the common value of $\underline{C}$ and $\overline{C}$.
This $C$, which we refer to simply as the \emph{capacity} of the DFSC,
can be expressed alternatively. If we assign a probability distribution to the
initial state, so that $s_0$ becomes a random variable,
then $C = \lim_{n\to\infty} C_n$, where
\begin{equation}
C_n = \frac1n \, \max_{Q^n(\bfx^n)} I(\bfx^n;\bfy^n \mid s_0).
\label{Cn_def}
\end{equation}
Clearly, $\underline{C}_n \le C_n \le \overline{C}_n$ for all $n$,
so that $C$, as defined above, is indeed the common value of
$\underline{C}$ and $\overline{C}$. Note that this is independent
of the choice of the probability distribution on $s_0$.

A further simplification to the expression for capacity is possible.
Since $|I(\bfx^n;\bfy^n) - I(\bfx^n;\bfy^n \mid s_0)| \le \log_2|\cS|$
(see, for example, \cite[Appendix~4A, Lemma~1]{Gallager1968}),
we in fact have
\begin{equation}
C = \lim_{n\to\infty} \frac1n \, \max_{Q^n(\bfx^n)} I(\bfx^n;\bfy^n).
\label{cap_eq}
\end{equation}

The capacity of a DFSC is difficult to compute in general. A useful lower bound
that is sometimes easier to compute (or at least estimate) is the so-called
\emph{symmetric information rate} (SIR) of the DFSC:
\begin{equation}\label{SIR_def}
R = \lim_{n\to\infty} \frac1n \, I(\bfx^n;\bfy^n),
\end{equation}
where the input sequence $\bfx$ is an i.i.d.\ Bernoulli($\nicefrac12$) 
random sequence.

\subsection{First results}
It is easy to see that the grains channel is a DFSC,
where the $n$th state $s_n$ is the pair $(u_n,x_n)$,
which takes values in the finite set ${\cS} =
\{(0,0),(0,1),(1,0),(1,1)\}$. Again, for completeness,
we assume an initial state $s_0$ that takes values
in $\cS$.\footnote{To be strictly faithful to the granular medium
we are modeling, we should restrict $s_0$ to take values only in
$\{(1,0),(1,1)\}$, so that $u_0 = 1$. This would imply $u_1 =0$,
meaning that no length-2 grain ends at the first bit cell of the medium,
corresponding to physical reality. But this makes no difference
to the asymptotics of the channel, and in particular, to the channel capacity.}

\begin{proposition}
The grains channel is indecomposable for $p < 1$.
\label{grains_indecomp_prop}
\end{proposition}
\begin{IEEEproof}
We must check that the condition in (\ref{indecomp_cond}) holds.
We take $n=1$ and $s_1 = (0,x_1)$.
Then, $\min_{s_0} q(s_1 \mid x_1,s_0)
= \min_{j \in \{0,1\}} P(u_1 = 0 \mid u_0 = j) = 1-p > 0$.
\end{IEEEproof}

\medskip

As a consequence of the above proposition, the equality 
$\underline{C} = \overline{C}$ holds for the grains channel when $p < 1$.
In fact, this equality also holds for the grains channel when $p = 1$,
as the following result shows.

\begin{proposition}
For the grains channel with $p=1$, we have
$\underline{C} = \overline{C} = \half$.
\label{Cg1_prop}
\end{proposition}
\begin{IEEEproof}
We have, with probability 1,
  \begin{align*}
\u &= u_1,u_2,u_3,u_4,u_5,u_6,\ldots\ \\&= \ \begin{cases}
0,1,0,1,0,1,\ldots & \text{ if } u_0 = 1 \\
1,0,1,0,1,0,\ldots & \text{ if } u_0 = 0.
\end{cases}
\end{align*}
Thus, once the initial state $s_0 = (u_0,x_0)$ is fixed, the output $\y$ of
the grains channel is a deterministic function of the input $\x$:
    \begin{align*}
\y &= y_1,y_2,y_3,y_4,y_5,y_6,\ldots\ \\&= \ \begin{cases}
x_1,x_1,x_3,x_3,x_5,x_5,\ldots & \text{ if } s_0 = (1,x_0) \\
x_0,x_2,x_2,x_4,x_4,x_6,\ldots & \text{ if } s_0 = (0,x_0).
\end{cases}
    \end{align*}
Therefore, for any fixed $s \in \cS$, we have $H(\y^n \mid \x^n, s_0 = s) = 0$,
and hence, $I(\x^n;\y^n \mid s_0 = s) = H(\y^n \mid s_0 = s)$.
If $\x^n$ is a sequence of i.i.d.\ Bernoulli($\nicefrac12$)
random variables, then $\min_{s \in \cS} H(\y^n \mid s_0 = s) =
 H(\y^n \mid s_0 = (0,x_0)) = \lfloor n/2 \rfloor$.
It follows that $\underline{C}_n \ge \frac{\lfloor n/2 \rfloor}{n}$,
so that $\underline{C} \ge 1/2$.
On the other hand, for any input distribution $Q^n(\x^n)$, and any
$s \in \cS$, we have $H(\y^n \mid s_0 = s) \le \lceil n/2 \rceil$.
Consequently, $\overline{C}_n \le \frac{\lceil n/2 \rceil}{n}$, and hence,
$\overline{C} \le 1/2$. We conclude that $\underline{C} = \overline{C} = 1/2$.
\end{IEEEproof}

\medskip

In view of the two propositions above, the capacity of the grains channel 
is defined by (\ref{cap_eq}). From here onward, we denote this capacity 
by $C^\g$, and use the notation $C^\g(p)$ when the dependence on $p$
needs to be emphasized. It is difficult to compute the capacity $C^\g$
exactly, so we will provide useful upper and lower bounds. We note here
for future reference the trivial bound obtained from 
Proposition~\ref{Cg1_prop}:
\begin{equation}
C^\g(p) \ge C^\g(1) = \half.
\label{cap_triv_bound}
\end{equation}

\subsection{Upper Bound: BINAEras}

Consider a binary-input channel similar to the
binary erasure channel, except that erasures in consecutive positions
are not allowed. Formally, this is a channel with a binary input sequence
$\bfx = x_1,x_2,x_3,\ldots$, with $x_i \in \{0,1\}$ for all $i$,
and a ternary output sequence $\y = y_1,y_2,y_3,\ldots$, with
$y_i \in \{0,1,\e\}$ for all $i$, where $\e$ is an erasure symbol.
The input-output relationship is determined by a binary sequence
$\u = u_1,u_2,u_3,\ldots$, which is a first-order Markov chain,
independent of the input sequence $\x$, with transition probabilities
$P(u_i | u_{i-1})$ as in (\ref{transit_probs}). We then have
\begin{equation}
y_i = \begin{cases}
x_i & \text{ if } u_i = 0 \\
\e & \text{ if } u_i = 1 \\
\end{cases}
\label{BINAEras_def}
\end{equation}
Since $P(u_i = 1 \mid u_{i-1} = 1) = 0,$ adjacent erasures do not occur, so
we term this channel the binary-input no-adjacent-erasures (BINAEras) channel.
To describe the channel completely, we define an initial state $z_0$
taking values in $\{0,\e\}$.

The BINAEras channel is a DFSC for which $\underline{C} = \overline{C}$ holds,
and its capacity, which we denote by $C^{\e}(p)$, can be computed explicitly.
\begin{theorem}
For the BINAEras channel with parameter $p \in [0,1]$,
we have $\underline{C} = \overline{C} = C^{\e}(p)\triangleq
  \frac{1}{1+p}.$
\label{Ce_theorem}
\end{theorem}
Intuitively, the average erasure probability of a symbol equals $\tilde p=\frac p
{1+p},$ and the capacity $C^{\e}(p)$ equals $1-\tilde p.$
A formal proof is given in Appendix~A.

We claim that the grains channel is a stochastically degraded BINAEras channel.
Indeed, the grains channel is obtained by cascading the BINAEras
channel with a ternary-input channel defined as follows: the input sequence
$\y = y_1,y_2,y_3,\ldots$, $y_i \in \{0,1,\e\}$, is transformed to the
output sequence $\y' = y'_1,y'_2,y'_3,\ldots$ according to the rule
\begin{equation}
y'_i =
\begin{cases} y_i & \text{ if } y_i \neq \e \\
y_{i-1} & \text{ if } y_i = \e
\end{cases}
\label{yi_y'i_eq}
\end{equation}
To cover the case when $y_1 = \e$, we set $y'_1$ equal to some arbitrary
$y_0 \in \{0,1\}$. It is straightforward to verify, via
(\ref{BINAEras_def}), (\ref{yi_y'i_eq}) and the fact that
$P(u_i = 1 \mid u_{i-1} = 1) = 0$,
that the cascade of the BINAEras channel with the above channel
has an input-output mapping $x_i \mapsto y'_i$ given by 
the equation obtained by replacing $y_i$ with $y_i'$ in (\ref{xi_yi_eq2}).
This immediately leads to the following theorem.
\vspace*{.05in}
\begin{theorem}
For $p \in [0,1]$, we have $C^\g(p) \le C^{\e}(p) = \frac{1}{1+p}$.
\label{Cg_upbnd}
\end{theorem}
{\em Remark:} We remark that any code that corrects $t$ nonadjacent
substitution errors (bit flips) also corrects $t$ grain errors.
It is therefore tempting to bound the capacity of the grains channel by the
capacity of the binary channel with {\em nonadjacent errors}.
Such a channel is defined similarly to the BINAEras channel: the channel noise
is controlled by a first-order Markov channel $\bfu$ (\ref{transit_probs}),
and $y_i=x_i \oplus u_i$ for all $i\ge 1.$
The capacity of this channel is computed as in the BINAEras case and equals
$1-\h(p)/(1+p),$ where $\h(p)$ denotes the binary entropy function.
However, a closer examination convinces one that this quantity does not
provide a valid lower bound for $C^\g(p)$.

\subsection{Lower Bound: The Symmetric Information Rate}\label{sec:lb}

In this section, we derive an exact expression for the SIR of the grains
channel, which gives a lower bound on the capacity of the channel.
In accordance with the definition of SIR (\ref{SIR_def}),
assume that $\x$ is an i.i.d.\ Bernoulli($\nicefrac12$) random sequence.
With this assumption, the state sequence $\s$ is a first-order Markov chain.
Also, each output symbol $y_n$ is easily verified to be
a Bernoulli($\nicefrac12$) random variable
(but $y_n$ is not independent of $y_{n-1}$).

We also assume that the initial state $s_0$ is a random variable distributed
according to the stationary distribution of the Markov chain, so that
the sequence $\s$ is a stationary Markov chain. It follows that the output
sequence $\y$ is a stationary random sequence, so that the entropy rate
$H(Y) := \lim_{n\to\infty} \frac1n \, H(\y^n)$ exists.
It is also worth noting here that the initial distribution assumed on
$s_0$ causes the Markov chain $\u$ to be stationary as well. In particular,
the random variables $u_i$, $i \ge 0$, all have the stationary distribution
given by $P(u_i = 0) = \frac{1}{1+p}$ and $P(u_i = 1) = \frac{p}{1+p}$.

We have
    \begin{align}
   R^\g&=\lim_{n\to\infty} \frac1n \, I(\x^n; \y^n)\label{eq:rg}\\[.05in]
 I(\x^n; \y^n)
&=  H(\y^n) - H(\y^n|\x^n)
=  H(\y^n) - H(\z^n| \x^n)
\label{Rn_start_eq}
\end{align}
 As noted above,
$H(Y) = \lim_{n\to\infty} \frac1n \, H(\y^n)$ exists. In fact, we can give
an exact expression for $H(Y)$ in terms of an infinite series.
\begin{proposition}\label{HY_prop}
The entropy rate of the output process of the grains channel is given by
$$
H(Y)
 = \frac{1}{2(1+p)}\sum_{j=2}^\infty \h(\beta_j) \prod_{k=2}^{j-1} (1-\beta_k),
$$
where
  $$
  \beta_j :=  \Pr[y_{j+1}  = 1\mid y_j = y_{j-1} = \cdots = y_2 = 0, y_1 = 1]
  $$
is given by the following recursion: $\beta_2 = \frac12 (1-p)$,
and for $j \ge 3$,
\begin{equation}
\beta_j
  = \frac12\left(\frac{1-(1+p) \beta_{j-1}}{1-\beta_{j-1}}\right).
\label{beta_rec}
\end{equation}
\end{proposition}
The lengthy proof of this proposition is given in Appendix~B.

\vspace*{.1in}{\em Remark:} The following
explicit expression for $\beta_j,j\ge 2$ can be proved by induction from (\ref{beta_rec}): 
\begin{equation}
    \beta_j=\frac{2({(\vartheta_-)}^j-{(\vartheta_+)}^j)}
	{(3+B+p){(\vartheta_-)}^j-(3-B+p){(\vartheta_+)}^j}
\label{beta_expr}
\end{equation}   
where $\vartheta_{\pm}=1-\frac{1\mp B}p$ and $B=\sqrt{p^2+6p+1}.$


\bigskip

Our next result shows that $\lim_{n\to\infty} H(\z^n \mid \x^n)$ also exists, 
and gives an exact expression for it, again in terms of an infinite series.
Appendix~B contains a proof of this result. 

\begin{proposition}
When $\x$ is an i.i.d.\ uniform Bernoulli sequence, we have
$$
\lim_{n \to \infty} H(\z^n \mid \x^n) 
 = \frac{1+p/2}{1+p} \,
\sum_{j=2}^\infty 2^{-j}\h\Big(\frac{1-(-p)^j}{1+p}\Big).
$$
\label{lim_Hzx_prop} 
\end{proposition}

Together, (\ref{eq:rg}), (\ref{Rn_start_eq}), 
and Propositions~\ref{HY_prop} and \ref{lim_Hzx_prop}
provide an {\em exact expression} for the SIR of the grains channel.
This, along with the trivial bound (\ref{cap_triv_bound}), yields 
the following lower bound on the capacity $C^\g$.
\begin{theorem}\label{Cg_lobnd}
The capacity $C^\g(p)\ge \max(\nicefrac12,R^\g(p)),$ 
where $R^\g(p)$ is the SIR of the grains channel and
is given by the following expression:
  \begin{multline*}
  R^\g(p)=\frac1{2(1+p)}\sum_{j=2}^\infty \biggl\{\h(\beta_j) \prod_{k=2}^{j-1} (1-\beta_k)
\\-\frac{2+p}{2^j} \, \h\Big(\frac{1-(-p)^j}{1+p}\Big)\biggr\}.
  \end{multline*}
with $\beta_j$ as in (\ref{beta_rec}) or (\ref{beta_expr}).
\end{theorem}

\medskip

\begin{figure}\centering
\subfigure[Bounds on $C^\g(p)$. The gray area
shows the gap between the lower bound of Theorem~\ref{Cg_lobnd}
and the upper bound of Theorem~\ref{Cg_upbnd}.]
{\includegraphics[scale=.9]{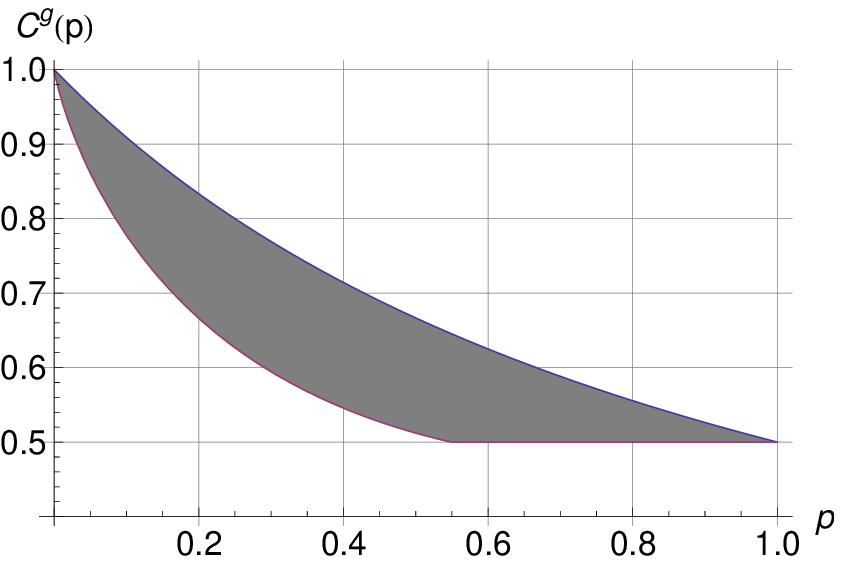}
\label{fig:subfig2}}
\vspace*{.1in}\subfigure[The symmetric information rate $R^\g(p)$.]{
\includegraphics[scale=.9]{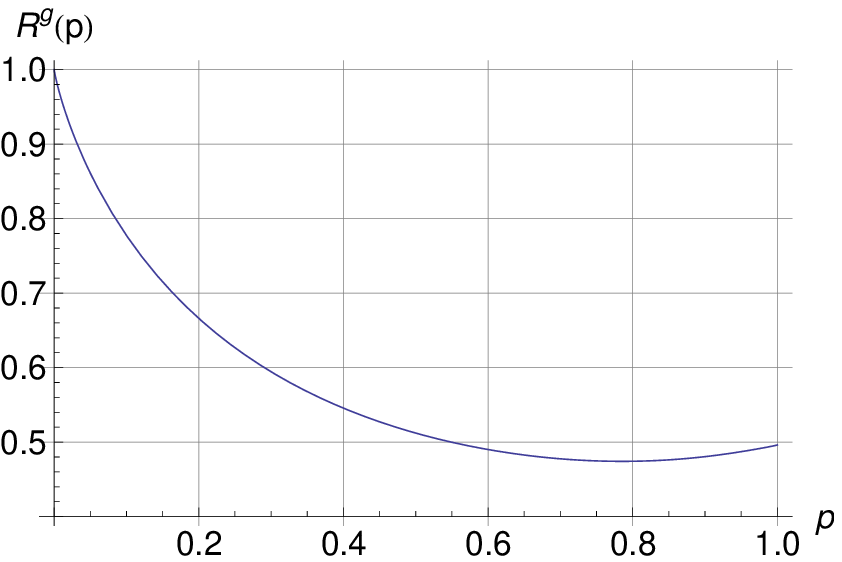}
\label{fig:subfig3}
}
\caption{Plots of the upper and lower bounds on the capacity 
of the grains channel $C^\g(p)$, and the SIR of the grains channel
$R^\g(p0$, as functions
of $p$. } \label{fig3}
\end{figure}

\remove{\begin{figure}\centering
\includegraphics[scale=.8]{C3.eps}
\caption{Plots of the upper and lower bounds on the capacity 
of the grains channel $C^g(p)$, and the SIR of the grains channel $R^g(p)$.The gray area
shows the gap between the lower bound of Theorem~\ref{Cg_lobnd}
and the upper bound of Theorem~\ref{Cg_upbnd}.}
\label{fig3}
\end{figure}}

In Figure~\ref{fig3}, we plot the upper and lower bounds on $C^\g(p)$ 
stated in Theorems~\ref{Cg_upbnd} and \ref{Cg_lobnd} as well as the value of $R^g(p)$
from Theorem~\ref{Cg_lobnd}. Observe that the SIR is 
a strict lower bound on the capacity, at least for $0.56 \le p < 1$,
when $R^\g(p) < \half$.

The plots are obtained by numerically evaluating $R^\g(p)$
by truncating its infinite series at some large value of $j$. We give here
a somewhat crude, but useful, estimate of the error in truncating this series 
at some index $j = J$, with $J \ge 2$. Define the partial sums
\begin{eqnarray}
S_J 
 &=& \frac{1+p/2}{1+p} \, \sum_{j=2}^J 2^{-j}\h\Big(\frac{1-(-p)^j}{1+p}\Big)
\label{SJ_def} \\
T_J 
 &=& \frac{1}{2(1+p)}\sum_{j=2}^J \h(\beta_j) \prod_{k=2}^{j-1} (1-\beta_k)
\label{TJ_def}
\end{eqnarray}
and note that the $J$th partial sum of the $R^\g(p)$ series is precisely
$T_J - S_J$. 

\begin{proposition}
The error $|R^g(p) - (T_J - S_J)|$ in truncating the $R^\g(p)$ series 
at an index $j = J$, with $J \ge 2$, is at most 
$$
\frac{1}{1+p} \, \left[(1+p/2) \, 2^{-J} + 2^{-\lfloor{(J+1)/2}\rfloor}\right].
$$
In particular, for any $p \in [0,1]$, the truncation error is at most 
$2^{-J} + 2^{-\lfloor{(J+1)/2}\rfloor}$.
\label{Rg_err_prop}
\end{proposition}

We defer the proof to Appendix~B.

The plot of $R^g(p)$ in Figure~\ref{fig3}(a) was generated using $J = 15$ terms 
of the infinite series, so the plotted curve is within
0.004 of the true $R^\g$ curve for all $p.$

\remove{
In the following we derive bounds on $H(Y)$ that can provide easier-to-compute
bounds on $C^\g$
\begin{lemma}
    \begin{multline}\label{eq:HY_lobnd}
 \frac{p}{1+p} + \frac{1}{1+p} \h\Big(\frac{1+p}{2}\Big)\le H(Y)
   \\\le \frac{1}{2(1+p)} \h\Big(\frac{1-p}{2}\Big)
     + \frac{1+2p}{2(1+p)} \h \Big(\frac{1+p}{2(1+2p)}\Big)
    \end{multline}
\end{lemma}
\begin{IEEEproof}
We will prove the lower bound because it is directly relevant to our main problem
of bounding $C^\g,$ and omit the proof of the upper bound.

We have $\lim_{n\to \infty} \frac{1}{n} \H(\bfy^n \mid s_0) = \lim_{n\to\infty}
\frac1n \, \sum_{i=1}^n H(y_i \mid \bfy^{i-1}, s_0)
= \lim_{i \to\infty} H(y_i \mid \bfy^{i-1},s_0)$.
Since conditioning reduces entropy,
we have $H(y_i \mid \bfy^{i-1}) \ge H(y_i \mid \bfy^{i-1},u_{i-1})$.
Now, (\ref{xi_yi_eq1}) implies that
$y_i$ is conditionally independent of $y_1,\ldots,y_{i-2}$ given $u_{i-1}$.
Hence, $$H(y_i \mid \bfy^{i-1},u_{i-1}) = H(y_i \mid y_{i-1},u_{i-1}).$$
We will show that
   $$
H(y_i \mid y_{i-1},u_{i-1})
= \frac{p}{1+p} + \frac{1}{1+p} \h\Big(\frac{1+p}{2}\Big),
   $$
which will prove the lemma. This is just a somewhat tedious computation.

We start with the identity
\begin{align}
H(y_i \mid &y_{i-1},u_{i-1})
= \sum_{(a,b) \in \{0,1\}^2} H(y_i \mid y_{i-1} = a, u_{i-1} = b)\nonumber\\
&\quad\times\Pr[y_{i-1} = a , u_{i-1} = b].
\label{lem10_eq1}
\end{align}
Given $u_{i-1} = 1$, we have (with probability 1) $u_i = 0$,
so that $y_i = x_i$. Thus,
\begin{align*}
H(y_i \mid y_{i-1} = a, u_{i-1} = 1)
&= H(x_i \mid y_{i-1} = a, u_{i-1} = 1)\\
& = H(x_i) = 1.
\end{align*}
Next, given $u_{i-1} = 0$, we have $y_{i-1} = x_{i-1}$, so that
$$
H(y_i \mid y_{i-1} = a,u_{i-1} = 0) = H(y_i \mid x_{i-1}=a,u_{i-1} = 0),
$$
and furthermore,
\begin{align*}
\Pr[&y_{i-1} = a, u_{i-1} = 0]\\
&= \Pr[u_{i-1} = 0] \Pr[y_{i-1} = a \mid u_{i-1} = 0]  \\
&= \Pr[u_{i-1} = 0] \Pr[x_{i-1} = a \mid u_{i-1} = 0]\\
&= \half \Pr[u_{i-1} = 0].
\end{align*}
Thus, the right-hand side of (\ref{lem10_eq1}) simplifies to
\begin{align}
\Pr[u_{i-1} = 1]
 &+ \half \Pr[u_{i-1} = 0] \nonumber\\
      &\quad\times \sum_{a \in \{0,1\}} H(y_i \mid x_{i-1}=a,u_{i-1} = 0).
\label{lem10_eq2}
\end{align}

To evaluate $H(y_i | x_{i-1}=0, u_{i-1} = 0)$, we compute
   \begin{align*}
      \Pr[y_i = &0 | x_{i-1} = 0, u_{i-1} = 0]\\
   &=
\sum_{j \in \{0,1\}} \Pr[y_i = 0 | x_{i-1} = 0, u_{i-1} = 0, u_i = j]\\
   &\quad\times \Pr[u_i = j | u_{i-1} = 0].
   \end{align*}
Note that, given $u_i = 0$, we have $y_i = x_i$, and hence
   \begin{align*}
\Pr[y_i = &0 \mid x_{i-1} = 0, u_{i-1} = 0, u_i = 0]\\
&= \Pr[x_i = 0 \mid x_{i-1} = 0, u_{i-1} = 0, u_i = 0] \\
&= \Pr[x_i = 0] = 1/2.
   \end{align*}
On the other hand, given $u_i = 1$, we have $y_i = x_{i-1}$, from which
we obtain
$$\Pr[y_i = 0 \mid x_{i-1} = 0, u_{i-1} = 0, u_i = 1] = 1.$$
Hence, $\Pr[y_i = 0 \mid x_{i-1} = 0, u_{i-1} = 0] = \frac{1+p}{2}$.
An analogous argument shows that
$\Pr[y_i = 1 \mid x_{i-1} = 1, u_{i-1} = 0] = \frac{1+p}{2}$.
Thus, (\ref{lem10_eq2}) evaluates to
\begin{equation}
\Pr[u_{i-1} = 1]
 + \Pr[u_{i-1} = 0] \, \h\Big(\frac{1+p}{2}\Big).
\label{lem10_eq3}
\end{equation}
The above expression is precisely the lower bound in
the statement of the lemma.
\end{IEEEproof}

We thus obtain the following lower bound on the capacity of the grains channel:
\begin{align*}
C^\g(p) \ge& \frac{p}{1+p}+\frac{1}{1+p}\h\Big(\frac{1+p}{2}\Big) \\
&-
\frac{1+p/2}{1+p} \,
\sum_{j=2}^\infty \h\Big(\frac{1-(-p)^j}{1+p}\Big) 2^{-j}.
\end{align*}
}

\subsection{Zero-Error Capacity}\label{sec:zero-error}

We end with a few remarks on the zero-error capacity of the grains channel.
We are interested in the maximum zero-error information rate,
$R_0(n)$, achievable over the grains channel with parameter $p \in [0,1]$
and input $\x^n$.  The case when $p=0$ is trivial
(the channel introduces no errors), so we consider $p > 0$.

The zero-error analysis depends on the initial state $s_0$ of the channel.
Suppose that $s_0$ is such that $\Pr[u_1 = 1] > 0$. Then, the state sequence
$\u^n = 1,0,1,0,\ldots,(n \!\! \mod 2)$ is realized with some
positive probability. Corresponding to this state sequence, we have
$\y^n = x_0,x_2,x_2,x_4,\ldots,x_{2\lfloor{n/2}\rfloor}$. Thus, at most
$\lfloor n/2 \rfloor$ bits can be transmitted without error across this
realization of the channel. Hence, $R_0(n) \le \frac1n \, \lfloor n/2 \rfloor$.
This zero-error information rate can actually be achieved. Consider
the binary length-$n$ code $\cR_n$ defined in (\ref{eq:Rn})
which has $2^{\lfloor n/2 \rfloor}$ codewords. When a codeword
from $\cR_n$ is sent across \emph{any} realization of the grains channel,
the bits at even coordinates remain unchanged. Thus, $\lfloor n/2 \rfloor$
bits of information can be transmitted without error, which proves that
$R_0(n) = \frac1n \, \lfloor n/2 \rfloor$.

On the other hand, suppose that the initial state $s_0$ is such that
$\Pr[u_1 = 1] = 0$. Then, the worst-case channel realization is caused
by the state sequence $\u^n = 0,1,0,1,\ldots,(1+n \!\! \mod 2)$. In this
case, the channel is such that the first coordinate of the input sequence
is always received without error at the output. A slight modification of
the preceding argument now shows that $R_0(n) = \frac1n \, \lceil n/2 \rceil$.

We have thus proved the following result.

\begin{proposition}
Consider a grains channel with parameter $p > 0$.
If the initial state $s_0$ is such that $\Pr[u_1 = 1] > 0$,
then $R_0(n) = \frac1n \, \lfloor n/2 \rfloor$; otherwise,
$R_0(n) = \frac1n \, \lceil n/2 \rceil$.

In any case, the zero-error capacity of the channel is
$C_0 = \lim_{n\to\infty} R_0(n) = \half$.
\label{zero_error_prop}
\end{proposition}

\section*{Appendix~A: Proof of Theorem~\ref{Ce_theorem}}
Observe first that the BINAEras channel is indecomposable for $p < 1$.
Indeed, for this channel, the condition in (\ref{indecomp_cond}) reduces
to showing that for some fixed $n$, there exists a choice for $u_n$ such that
$\min_{u_0} P(u_n | u_0) > 0$. This condition clearly holds for $n=1$
and $u_1 = 0$: $\min_{j \in \{0,1\}} P(u_1 = 0 \mid u_0 = j) = 1-p > 0$,
provided $p < 1$. We deal with the indecomposable case in this appendix;
when $p=1$, the proof for $\underline{C} = \overline{C} = \half$ follows,
\emph{mutatis mutandis}, the proof of Proposition~\ref{Cg1_prop}.

When the channel is indecomposable,
we have $\underline{C} = \overline{C} = C$. We will show that
$C = \frac{1}{1+p}$. Choose the distribution on $u_0$ to be
the stationary distribution of the Markov process $\u$, so that
$P(u_0 = 0) = \frac{1}{1+p}$ and $P(u_0 = 1) = \frac{p}{1+p}$.
Consequently, $\u$ is a stationary process, and in particular,
for all $i \ge 1$, we have $P(u_i = 0) = \frac{1}{1+p}$
and $P(u_i = 1) = \frac{p}{1+p}$.

Observe that
\begin{align*}
I(\x^n;\y^n \mid u_0) &= H(\y^n \mid u_0) - H(\y^n \mid \x^n,u_0)\\
&\stackrel{(a)}{=} H(\y^n \mid u_0) - H(\u^n\mid \x^n,u_0) \\
&\stackrel{(b)}{=} H(\y^n \mid u_0) - H(\u^n \mid u_0),
\end{align*}
with equality (a) above due to the fact that, given $\x^n$, the sequences
$\y^n$ and $\u^n$ uniquely determine each other, and equality (b)
because $\u^n$ is independent of $\x^n$. Furthermore, since $\u$ is
a stationary first-order Markov process, we have $H(\u^n \mid u_0)
= \sum_{n=1}^n H(u_n \mid u_{n-1}) = n H(u_1 \mid u_0) = n \frac{\h(p)}{1+p}$.
Hence,
\begin{equation}
C_n = n^{-1} \max_{Q^n(\x^n)} H(\y^n \mid u_0) - \frac{\h(p)}{1+p}.
\label{CN_eq}
\end{equation}

Now, $H(\y^n \mid u_0) = \sum_{i=1}^{n} H(y_{i} \mid \y^{i-1},u_0)$.
Since $\y^{i-1}$ completely determines $\u^{i-1}$, we have by
the data processing inequality \cite[Theorem~2.8.1]{CT06},
$$
H(y_i \mid \y^{i-1},u_0) \le H(y_i\mid\u^{i-1},u_0)
$$
We further have
   \begin{align*}
 H(&y_i\mid\u^{i-1},u_0) \le H(y_i \mid u_{i-1})\\
&= H(y_i \mid u_{i-1} = 0) \frac{p}{1+p} +
  H(y_i \mid u_{i-1} = 1) \frac{1}{1+p}
  \end{align*}
Given $u_{i-1} = 1$, $\y_i$ is a binary random variable
(since $u_i = 0$ with probability 1), and thus,
$H(y_i \mid u_{i-1} = 1) \le 1$.
On the other hand, we have $P(y_i = \e \mid u_{i-1} = 0)
= P(u_i = 1 \mid u_{i-1} = 0) = p$, and so the conditional entropy
$H(y_i \mid u_{i-1} = 0)$ is maximized when $P(y_i = 0 \mid u_{i-1} = 0)
= P(y_i = 1 \mid u_{i-1} = 0) = (1-p)/2$. This yields
$H(y_i \mid u_{i-1} = 1) \le \h(p) + 1-p$. Putting all the inequalities
together, we find that
    \begin{align*}
H(\y^n \mid u_0) &= \sum_{i=1}^{n} H(y_{i} \mid \y^{i-1},u_0)\\
  &\le n \Big( \frac{p}{1+p} + (\h(p) + 1-p)\frac{1}{1+p} \Big)\\
   &= n \Big(\frac{1 + \h(p)}{1 + p}\Big)
   \end{align*}
It is not difficult to check that the above in fact holds with equality when
the input sequence $\x^n$ is an i.i.d.\ sequence of Bernoulli($\nicefrac12$) random
variables. Thus,
$$
n^{-1} \max_{Q^n(\x^n)} H(\y^n \mid u_0) = \frac{1 + \h(p)}{1 + p}.
$$
Plugging this into (\ref{CN_eq}), we obtain that $C_n = \frac{1}{1+p}$
for all $n$, and hence, $C = \frac{1}{1+p}$.
%

\section*{Appendix~B: Proofs of Propositions~\ref{HY_prop}, \ref{lim_Hzx_prop}
and \ref{Rg_err_prop}}

\subsection*{B.1. Proof of Proposition~\ref{HY_prop}}
Since $\lim_{n\to\infty} \frac1n H(\y^n) = 
\lim_{i\to\infty} H(y_{i+1} \mid y^i)$, we need show that
the latter limit equals the expression in the statement of the proposition. 
We will work with the identity
$$
H(y_{i+1} \mid \y^i)
  = \sum_{\b \in \{0,1\}^i} H(y_{i+1} \mid \y^i = \b) \Pr[\y^i = \b].
$$
From the channel input-output relationship given by (\ref{xi_yi_eq2})
and the fact that the input $\x$ is an i.i.d.\ Bernoulli($\half$) sequence,
it is clear that $\Pr[\y^i = \b] = \Pr[\y^i = \bar{\b}]$, where
$\bar{\b} = \b+1^n$ is the sequence obtained by flipping each bit in $\b$.
It then also follows that
$H(y_{i+1} \mid \y^i = \b) = H(y_{i+1} \mid \y^i = \bar{\b})$,
since
$\Pr[y_{i+1} = 1 \mid \y^i = \b] = \Pr[y_{i+1} = 0 \mid \y^i = \bar{\b}]$.
Hence,
\begin{equation}
H(y_{i+1} \mid \y^i)
  = 2 \, \sum_{\b \in B} H(y_{i+1} \mid \y^i = \b) \Pr[\y^i = \b],
\label{Hy_eq1}
\end{equation}
where $B = \{(b_i,\ldots,b_1) \in \{0,1\}^i: b_i = 0\}$ is the set of
all binary length-$i$ sequences that have a 0 in the leftmost coordinate.

Fix $i \ge 2$. Define, for $2 \le j \le i$, the events
$$
 B_j = \{\y^i : (y_i,y_{i-1},\ldots,y_{i-j+1}) = 0^{j-1}1\},
$$
which, together with the event $\{\y^i = 0^i\}$, form a partition of $B$.
Here, $0^{j-1}1$ is shorthand for the $j$-tuple $(0,\ldots,0,1)$.
We record two facts about $B_j$. First,
\begin{eqnarray}
\Pr[\y^i \in B_j] &=& \Pr[(y_i,y_{i-1},\ldots,y_{i-j+1}) = 0^{j-1}1] \notag \\
&=& \Pr[(y_j,y_{j-1},\ldots,y_1) = 0^{j-1}1],
\label{Pr_Bj_eq}
\end{eqnarray}
the last equality stemming from the fact that $\y$ is stationary.
Second, by the following lemma,
\begin{equation}
H(y_{i+1} \mid \y^i = \b) = \h(\Pr[y_{i+1} = 1 \mid \y^i = \b])
\label{Hy_eq2}
\end{equation}
is invariant over $B_j$.

\begin{lemma}\label{Bj_lemma}
For $\b \in B_j$, $\Pr[y_{i+1} = 1 \mid \y^i = \b]$ equals
$$
\half \Pr[u_j = 0 \mid (y_{j-1},y_{j-2},\ldots,y_2) = 0^{j-2}, \,
              (u_1,x_1) = (0,0)].
$$
\end{lemma}
\begin{IEEEproof}
The proof relies upon the following claim :

\vspace{.05in}\begin{minipage}{.9\linewidth}
Suppose that $y_{k-1} = b$; then, with probability 1, we have
$y_k = \bar{b}$ if and only if $s_k := (u_k,x_k) = (0,\bar{b})$.
\end{minipage}

\vspace{.05in}
\nd Indeed, even without the assumption on $y_{k-1}$,
the ``if'' part holds trivially.
For the ``only if'' part, assume that $y_{k-1} = b$ and $y_k = \bar{b}$.
Note that if $u_k = 1$, then with probability 1,
we have $u_{k-1} = 0$. Hence, by way of (\ref{xi_yi_eq2}), we have
$y_k = x_{k-1} = y_{k-1}$. However, $y_{k-1} \neq y_k$ by assumption;
so we must have $u_k = 0$. Consequently, $y_k = x_k$, so that $x_k = \bar{b}$.

Consider any $\b \in B_j$. From the claim, we have
\begin{eqnarray*}
\Pr[y_{i+1} = 1| \y^i= \b]
&=& \Pr[(u_{i+1},x_{i+1}) = (0,1)| \y^i = \b] \\
&=& \half \Pr[u_{i+1} = 0 | \y^i = \b],
\end{eqnarray*}
where we have used the fact that $x_{i+1}$ is independent of $\y^i$.
Note that, in the event $\y^i = \b$, we have $y_{i-j+2} = 0$
and $y_{i-j+1} = 1$, so that by the claim again,
$$
\Pr[u_{i+1} = 0|\y^i = \b]
= \Pr[u_{i+1} = 0| \y^i = \b, s_{i-j+2} = (0,0)].
$$
Now, given the channel state $s_{i-j+2} = (0,0)$,
the random variables $u_{i+1}$, $y_i$, $y_{i-1},\ldots,y_{i-j+2}$ are
conditionally independent of the past output $\y^{i-j+1}$.
Furthermore, given $s_{i-j+2} = (0,0)$,
the random variable $y_{i-j+2}$ is uniquely determined: $y_{i-j+2} = 0$.
Hence,
\begin{eqnarray*}
\lefteqn{\Pr[u_{i+1} = 0 \mid \y^i = \b, s_{i-j+2} = 0]  =} \\
& &  \Pr[u_{i+1} = 0
        \mid (y_i,\ldots,y_{i-j+3}) = 0^{j-2},\, s_{i-j+2} = 0].
\end{eqnarray*}
Finally, by the joint stationarity of $\y$ and $\u$, the right-hand side
above is equal to
$$
\Pr[u_{j} = 0 \mid (y_{j-1},y_{j-2},\ldots,y_2) = 0^{j-2},\, s_1 = 0],
$$
which is what we needed to show.
\end{IEEEproof}

In the statement of Proposition~\ref{HY_prop}, we defined
$\beta_j = \Pr[y_{j+1} = 1 \mid (y_j,y_{j-1},\ldots,y_1) = 0^{j-1}1]$.
Note that if we set $i=j$ in Lemma~\ref{Bj_lemma}, we get
\begin{equation}
\beta_j =
\half \Pr[u_{j} = 0 \mid (y_{j-1},\ldots,y_2) = 0^{j-2},\,
(u_1,x_1) = (0,0)].
\label{beta_eq}
\end{equation}
From (\ref{Hy_eq1})--(\ref{beta_eq}), and Lemma~\ref{Bj_lemma}, we have
\begin{eqnarray}
H(y_{i+1} \mid \y^i)
 &=& 2 \sum_{j = 2}^i \h(\beta_j) \Pr[(y_j,\ldots,y_1) = 0^{j-1}1] \notag \\
 & &  + \ 2 H(y_{i+1} \mid \y^i = 0^i) \Pr[\y^i = 0^i]. \label{Hy_eq3}
\end{eqnarray}

The term at the end of the above expression vanishes as $i \to \infty$,
as we show below for completeness.

\begin{lemma}
$\displaystyle
\lim_{i\to\infty} H(y_{i+1} \mid \y^i = 0^i) \Pr[\y^i = 0^i] = 0$.
\label{HYproof_lemma2}
\end{lemma}
\begin{IEEEproof}
Since $0 \le H(y_{i+1} \mid \y^i = 0^i) \le 1$, it is enough to
show that $\Pr[\y^i = 0^i] = 0$ converges to 0. For this, observe
that for any $j$, if $y_j = 0$, then $(x_{j-1},x_j) \neq (1,1)$.
Hence, if $\y^i = 0^i$, then $(x_1,x_2) \neq (1,1)$, $(x_3,x_4) \neq (1,1)$,
and so on. Thus, $\Pr[\y^i = 0^i] \le (3/4)^{\lfloor i/2 \rfloor}$,
which suffices to prove the lemma.
\end{IEEEproof}

So, letting $i \to \infty$ in (\ref{Hy_eq3}), we obtain
\begin{equation}
H(Y) =
 2 \sum_{j = 2}^{\infty} \h(\beta_j) \Pr[(y_j,y_{j-1},\ldots,y_1) = 0^{j-1}1].
\label{HY_eq1}
\end{equation}

The proof of Proposition~\ref{HY_prop} will be complete once we prove
the next two lemmas.
\begin{lemma}\label{HYproof_lemma3}
For $j \ge 2$, we have
$$
\Pr[(y_j,y_{j-1},\ldots,y_1) = 0^{j-1}1]
   = \frac{1}{4(1+p)} \prod_{k=2}^{j-1} (1-\beta_k)
$$
\end{lemma}
\begin{IEEEproof}
From the definition of $\beta_j$, we readily obtain
\begin{eqnarray*}
\lefteqn{\Pr[(y_j,\ldots,y_1) = 0^{j-1}1] = } \\
& & \left[\prod_{k=2}^{j-1} (1-\beta_k) \right] \cdot \Pr[(y_2,y_1) = (0,1)].
\end{eqnarray*}
We must show that $\Pr[(y_2,y_1) = (0,1)] = \frac{1}{4(1+p)}$. 

We write 
\begin{eqnarray*}
\lefteqn{\Pr[(y_2,y_1) = (0,1)] = } \\
& & \sum_{(a,b) \in \{0,1\}^2}
   \Pr[(y_2,y_1) = (0,1) \mid (u_2,u_1) = (a,b)] \\
& & \ \ \ \ \ \ \ \ \ \ \ \ \ \times \ \Pr[(u_2,u_1) = (a,b)].
\end{eqnarray*}
Clearly, $\Pr[(u_2,u_1) = (1,1)] = 0$. Also,
$\Pr[(y_2,y_1) = (0,1) \mid (u_2,u_1) = (1,0)] = 0$, since,
given $(u_2,u_1) = (1,0)$ we must have $y_2 = x_1 = y_1$, by virtue of
(\ref{xi_yi_eq2}). Next, given $(u_2,u_1) = (0,0)$, we have
$(y_2,y_1) = (x_2,x_1)$, and since $(x_2,x_1)$ is independent of $(u_2,u_1)$,
we find that
\begin{eqnarray*}
\lefteqn{\Pr[(y_2,y_1) = (0,1) \mid (u_2,u_1) = (0,0)]} \\
& & \ \ \ \ \ \ \ \ \ \ \ \ = \ \Pr[(x_2,x_1) = (0,1)] \ = \ \nicefrac14.
\end{eqnarray*}
By a similar argument, $\Pr[(y_2,y_1) = (0,1) \mid (u_2,u_1) = (0,1)]
= \nicefrac14$. Hence,
$$
\Pr[(y_2,y_1) = (0,1)] = (\nicefrac14) \Pr[u_2 = 0] = \frac{1}{4(1+p)},
$$
as desired.
\end{IEEEproof}

\begin{lemma}
$\beta_2 = \frac12(1-p)$, and for $j \ge 3$, $\beta_j$ satisfies the recursion
in (\ref{beta_rec}).
\label{HYproof_lemma4}
\end{lemma}
\begin{IEEEproof}
From (\ref{beta_eq}), we have
\begin{eqnarray*}
\beta_2 &=& \half \Pr[u_2 = 0 \mid (u_1,x_1) = (0,0)] \\
&=& \half \Pr[u_2 = 0 \mid u_1 = 0] \ \  = \ \ \half(1-p).
\end{eqnarray*}

For convenience, define, for $j \ge 2$,
$E_j = \{(y_{j-1},y_{j-2},\ldots,y_2) = 0^{j-2},\, (u_1,x_1) = (0,0) \}$,
so that $\beta_j = (\half)\Pr[u_j = 0 \mid E_j] = (\half)(1 - \gamma_j)$,
where $\gamma_j := \Pr[u_j = 0 \mid E_j]$. We shall show that for $j \ge 3$,
\begin{equation}
\gamma_j = \frac{p(1-\gamma_{j-1})}{1+\gamma_{j-1}}.
\label{gamma_rec}
\end{equation}
which is equivalent to the recursion in (\ref{beta_rec}).

So, let $j \ge 3$ be fixed. We start with
\begin{eqnarray*}
\gamma_j &=& \sum_{b \in \{0,1\}} \Pr[u_j=1 \mid u_{j-1} = b] \,
  \Pr[u_{j-1} = b \mid E_j] \notag \\
&=& p \cdot \Pr[u_{j-1} = 0 \mid E_j] \notag \\ 
&=& p \cdot \Pr[u_{j-1} = 0 \mid y_{j-1} = 0, E_{j-1}] \notag \\
&=& p \cdot \frac{\Pr[y_{j-1} = 0 \mid u_{j-1} = 0, E_{j-1}] (1-\gamma_{j-1})}{\Pr[y_{j-1} = 0 \mid E_{j-1}]}
\end{eqnarray*}
where we have used $\Pr[u_{j-1} = 0 \mid E_{j-1}] = 1 - \gamma_{j-1}$
for the last equality.

Given $u_{j-1} = 0$, we have $y_{j-1} = x_{j-1}$, and since $x_{j-1}$
is independent of $u_{j-1}$ and $E_{j-1}$, the numerator in the last expression
above evaluates to $\half (1-\gamma_{j-1})$. Thus,
\begin{equation}
\gamma_j = p \cdot \frac{\half(1-\gamma_{j-1})}{\Pr[y_{j-1} = 0 \mid E_{j-1}]}
\label{lemma19_eq1}
\end{equation}
Turning to the denominator, we write $\Pr[y_{j-1} = 0 \mid E_{j-1}]$ as
\begin{eqnarray}
\lefteqn{\sum_{b\in\{0,1\}} \Pr[y_{j-1} = 0 \mid u_{j-1} = b, E_{j-1}] \Pr[u_{j-1} = b \mid E_{j-1}]} \notag \\
&=& \textstyle{\half}(1-\gamma_{j-1}) + \Pr[y_{j-1} = 0 \mid u_{j-1} = 1, E_{j-1}]\cdot \gamma_{j-1} \notag \\
\label{lemma19_eq2}
\end{eqnarray}
We claim that $\Pr[y_{j-1} = 0 \mid u_{j-1} = 1, E_{j-1}] = 1$. Indeed, given
$u_{j-1} = 1$, we have $y_{j-1} = x_{j-2}$. Furthermore,
we must have $u_{j-2} = 0$ with probability 1, so that $x_{j-2} = y_{j-2}$.
Thus, given $u_{j-1} = 1$, we must have $y_{j-1} = y_{j-2}$ with probability 1.
But note that the event $E_{j-1}$ implies $y_{j-2} = 0$: if $j = 3$, this
follows from $(u_1,x_1) = (0,0)$, and if $j \ge 4$, this is contained
within $(y_{j-2},\ldots,y_2) = 0^{j-3}$. Thus,
given $u_{j-1} = 1$ and $E_{j-1}$, we have $y_{j-1} = y_{j-2} = 0$ with
probability 1.

So, carrying on from (\ref{lemma19_eq2}), we get
$$
\Pr[y_{j-1} = 0 \mid E_{j-1}] =
\half(1-\gamma_{j-1}) + \gamma_{j-1} = \half(1 + \gamma_{j-1})
$$
Feeding this back into (\ref{lemma19_eq1}), we obtain
$$
\gamma_j
 = p \cdot \frac{\half(1-\gamma_{j-1})}{\half(1+\gamma_{j-1})}
$$
which is the desired recursion (\ref{gamma_rec}).
\end{IEEEproof}

This concludes the proof of Proposition~\ref{HY_prop}.

\subsection*{B.2. Proof of Proposition~\ref{lim_Hzx_prop}}
We break the proof into two parts. We first show that
\begin{equation}
\lim_{n\to\infty} \frac1n \, H(\z^n \mid \bfx^n) =
\sum_{j=2}^\infty 2^{-j} \, H(u_j \mid u_1)
\label{lim_Hzx_eq1}
\end{equation}
and subsequently, we prove that 
\begin{equation}
\sum_{j=2}^\infty 2^{-j} \, H(u_j| u_1) 
= \frac{1+p/2}{1+p} \,
\sum_{j=2}^\infty 2^{-j} \, \h\Big(\frac{1-(-p)^j}{1+p}\Big).
\label{lim_Hzx_eq2}
\end{equation}

To show (\ref{lim_Hzx_eq1}), we start with 
$$
H(\z^n \mid \bfx^n) = \sum_{i=1}^n H(z_i \mid z_1,\ldots,z_{i-1},\bfx^n).
$$
From (\ref{zi}), it is evident that
$z_i$ is independent of $x_j$ for $j > i$. Hence,
    $$
H(\z^n \mid \bfx^n) = \sum_{i=1}^n H(z_i \mid z_1,\ldots,z_{i-1},\bfx^i).
   $$
As a result, by the Ces{\`a}ro mean theorem,
    $$
    \lim_{n \to \infty} \frac1n \, H(\z^n \mid \bfx^n)
= \lim_{i \to \infty} H(z_i \mid z_1,\ldots,z_{i-1},\bfx^i),
    $$
provided the latter limit exists.

To evaluate $H(z_i \mid z_1,\ldots,z_{i-1},\bfx^i)$, we define
the events $A_0 = \{\bfx^i: x_i = x_{i-1}\}$,
$$
A_j = \{\bfx^i: x_i \neq x_{i-1} = \cdots = x_{i-j} \neq x_{i-j-1}\},
 \ 1 \le j \le i-2,
$$
and $A_{i-1} = \{\bfx^i: x_i \neq x_{i-1} = \cdots = x_1\}$.
These events partition the space $\{0,1\}^i$ to which $\bfx^i$ belongs.
Since $\x$ is an i.i.d.\ uniform Bernoulli sequence, 
we have $\Pr[\bfx^i \in A_j] = (\half)^{j+1}$ for $0 \le j \le i-2$,
and $\Pr[\bfx^i \in A_{i-1}] = (\half)^{i-1}$.

Now, if $\bfx^i \in A_0$, then by (\ref{zi}), we have $z_i = 0$. Consequently,
$H(z_i \mid z_1,\ldots,z_{i-1},\bfx^i \in A_0) = 0$.

If $\bfx^i \in A_j$ for some $j \in [1,i-2]$, then we have $z_i = u_i$,
$z_{i-1} = \cdots = z_{i-j+1} = 0$, and $z_{i-j} = u_{i-j}$. Thus,
\begin{eqnarray*}
\lefteqn{H(z_i \mid z_1,\ldots,z_{i-1},\bfx^i \in A_j)} \\
&=& H(u_i \mid z_1,\ldots,z_{i-j-1},u_{i-j}, \bfx^i \in A_j) \\
&\stackrel{\text{(a)}}{=}& H(u_i \mid u_{i-j})
\ \ \stackrel{\text{(b)}}{=} \ \ H(u_{j+1} \mid u_1).
\end{eqnarray*}
Equality (a) above is due to the fact that $\u$ is a
first-order Markov chain independent of $\bfx$,
while equality $(b)$ is a consequence of the stationarity of $\u$
(which is itself a consequence of the stationarity of the state sequence 
$\bfs$).

Finally, if $\bfx^i \in A_{i-1}$, then $z_i = u_i$ and
$z_{i-1} = \cdots = z_2 = 0$. Thus,
$$
H(z_i \mid z_1,\ldots,z_{i-1},\bfx^i \in A_{i-1}) = H(u_i \mid z_1).
$$
Therefore,
\begin{eqnarray*}
\lefteqn{H(z_i \mid z_1,\ldots,z_{i-1},\bfx^i)} \\
&=& \sum_{j=0}^{i-1} H(z_i \mid z_1,\ldots,z_{i-1},\bfx^i \in A_j)
\Pr[\bfx^i \in A_j] \\
&=& \sum_{j=1}^{i-2} H(u_{j+1} \mid u_1) \, 2^{-j-1} +
H(u_i \mid z_1) \, 2^{-i+1}.
\end{eqnarray*}
Letting $i \to \infty$, we obtain (\ref{lim_Hzx_eq1}).

It remains to prove (\ref{lim_Hzx_eq2}). For this, 
note first that $H(u_j \mid u_1) 
= H(u_j \mid u_1 = 0) \Pr[u_1 = 0] + H(u_j \mid u_1 = 1) \Pr[u_1 = 1]$. 
Furthermore, since $u_1 = 1$ implies $u_2 = 0$ with probability 1, 
we have, for all $j \ge 2$,
$$
H(u_j \mid u_1 = 1) = H(u_j \mid u_2 = 0) = H(u_{j-1} \mid u_1 = 0),
$$
the last equality following from the stationarity of $\u$.
Hence,
   \begin{eqnarray*}
   \sum_{j=2}^\infty 2^{-j} \, H(u_j \mid u_1 = 1) 
&=& \sum_{j=2}^\infty 2^{-j} \, H(u_{j-1} \mid u_1 = 0) \\
&=& \frac12 \sum_{j=2}^\infty 2^{-j} \, H(u_{j} \mid u_1 = 0) 
   \end{eqnarray*}
since $H(u_1 \mid u_1 = 0) = 0$.
Putting it all together, we find that
\begin{eqnarray*}
\lefteqn{\sum_{j=2}^\infty 2^{-j} \, H(u_j \mid u_1)} \\
&=& (\Pr[u_1 = 0] + \frac12 \Pr[u_1 = 1])
\sum_{j=2}^\infty 2^{-j} \, H(u_j \mid u_1 = 0) \\
&=& \frac{1+p/2}{1+p} \, \sum_{j=2}^\infty 2^{-j} \, H(u_j \mid u_1 = 0).
\end{eqnarray*}

Finally, observe that 
$H(u_j \mid u_1 = 0) = \h\big(\frac{1 - (-p)^j}{1+p}\big)$, as it can be shown
(for example, by induction)
that $\Pr(u_j = 0 \mid u_1 = 0) = \frac{1 - (-p)^j}{1+p}$ for all $j \ge 1$.
This proves (\ref{lim_Hzx_eq2}), and with this, 
the proof of Proposition~\ref{lim_Hzx_prop} is complete.

\subsection*{B.3. Proof of Proposition~\ref{Rg_err_prop}}

The error in truncating the $R^\g(p)$ series at the index $j = J$ is 
\begin{eqnarray}
\lefteqn{|R^\g(p) - (T_J-S_J)|} \notag \\
& \le & 
|H(Y) - T_J| + |\lim_{n\to\infty} H(\z^n \mid \x^n) -  S_J|. \label{Rg_err}
\end{eqnarray}
It is easy to bound the second term in (\ref{Rg_err}):
\begin{eqnarray}
\lefteqn{| \lim_{n \to \infty} H(\z^n \mid \x^n) -  S_J |} \notag \\
&=& \frac{1+p/2}{1+p} \, 
 	\sum_{j=J+1}^\infty 2^{-j}\h\Big(\frac{1-(-p)^j}{1+p}\Big) \notag \\
&\le & \frac{1+p/2}{1+p} \, \sum_{j=J+1}^\infty 2^{-j} \notag \\
&=&  \left(\frac{1+p/2}{1+p}\right) \, 2^{-J}. \label{Hzx_err}
\end{eqnarray}
Turning our attention to the first term in (\ref{Rg_err}), we see that
\begin{eqnarray}
|H(Y) - T_J| 
&=& \frac{1}{2(1+p)}
	\sum_{j=J+1}^\infty \h(\beta_j) \prod_{k=2}^{j-1} (1-\beta_k)\notag \\
&\le & \frac{1}{2(1+p)}
	\sum_{j=J+1}^\infty \prod_{k=2}^{j-1} (1-\beta_k). \label{HY_err_eq1}
\end{eqnarray}
Now, from the recursion (\ref{beta_rec}), we readily get for $k \ge 3$,
$$
1 - \beta_k = \frac12\left(\frac{1-(1-p) \beta_{k-1}}{1-\beta_{k-1}}\right),
$$
and hence,
$$
(1-\beta_k)(1-\beta_{k-1}) = \frac12 [1 - (1-p) \beta_{k-1}] \leq \frac12.
$$
Consequently, if $j = 2m$ for some $m \geq 1$, then
$$
\prod_{k=2}^{j-1} (1-\beta_k) = 
\prod_{k=1}^{m-1} (1-\beta_{2k+1})(1-\beta_{2k}) \leq (\half)^{m-1},
$$
and if $j = 2m+1$ for some $m \geq 1$, then
$$
\prod_{k=2}^{j-1} (1-\beta_k) \le 
\prod_{k=1}^{m-1} (1-\beta_{2k+1})(1-\beta_{2k}) \leq (\half)^{m-1}.
$$
Upon replacing the bound in (\ref{HY_err_eq1}) by the looser 
$$
\frac{1}{2(1+p)} \sum_{j=2\lfloor{(J+1)/2}\rfloor}^\infty 
  \prod_{k=2}^{j-1} (1-\beta_k) 
$$
so that the summation starts at an even index $j$,
routine algebraic manipulations now yield
\begin{eqnarray*}
|H(Y) - T_J|
&\le& \frac{1}{1+p} \, \sum_{m=\lfloor{(J+1)/2}\rfloor}^\infty (\half)^{m-1} \\
&=& \left(\frac{1}{1+p}\right) 2^{-\lfloor{(J+1)/2}\rfloor}.
\end{eqnarray*}
Plugging this and (\ref{Hzx_err}) into (\ref{Rg_err}), we obtain 
Proposition~\ref{Rg_err_prop}.

\vspace*{.1in}\noindent{\sc Acknowledgement:}
Navin Kashyap would like to thank Bane Vasic for bringing to his
attention the problem considered in this work.
The authors would like to thank Haim Permuter for helpful discussions.

\end{document}